\documentstyle[aps,prd,epsfig]{revtex}
\begin{document}
\draft

\title{Quark structure of $f_0(980)$ from the radiative decays
$\phi(1020)\to\gamma f_0(980)$, $\gamma\eta$, $\gamma\eta'$,
$\gamma\pi^0$ and $f_0(980)\to\gamma\gamma$}
\author{A. V. Anisovich, V. V. Anisovich, V. A. Nikonov}
\address{St.Petersburg Nuclear Physics Institute, Gatchina, 188300,
Russia}
\date{20.03.2001}
\maketitle

\begin{abstract}
Partial widths of the radiative decays $\phi(1020)\to\gamma f_0(980)$,
$\gamma\eta$, $\gamma\eta'$, $\gamma\pi^0$ and
$f_0(980)\to\gamma\gamma$ are calculated assuming all mesons under
consideration to be $q\bar q$ states: $\phi(1020)$ is dominantly an
$s\bar s$ state ($n\bar n$ component $\lesssim 1\%$), $\eta$, $\eta'$
and $\pi^0$ are standard $q\bar q$ states, $\eta = n\bar n\cos\theta -
s\bar s\sin\theta$ and $\eta'=n\bar n\sin\theta + s\bar s\cos\theta$
with $\theta\simeq 37^\circ$, and $f_0(980)$ is the $q\bar q$ meson
with the flavour wave function $n\bar n\cos\varphi + s\bar s
\sin\varphi$. Calculated partial widths for the decays
$\phi(1020)\to\gamma \eta$, $\gamma\eta'$, $\gamma\pi^0$ are in a
reasonable agreement with experiment. The measured value of the
branching ratio $BR(\phi\to\gamma f_0(980))$ requires $25^\circ\le
|\varphi|\le 90^\circ$; for the decay $f_0(980)\to\gamma\gamma$ the
agreement with data is reached at either $77^\circ\le\varphi\le
93^\circ$ or $(-54^\circ)\le\varphi\le (-38^\circ)$. Simultaneous
analysis of the decays $\phi(1020)\to\gamma f_0(980)$ and
$f_0(980)\to\gamma\gamma$ provides arguments in favour of the solution
with negative mixing angle $\varphi =-48^\circ\pm 6^\circ$.
\end{abstract}

\pacs{13.40.Hq, 12.38.-t, 14.40.-n}

\section{Introduction}

The problem of the $f_0(980)$ structure is actively discussed
during the last decade, see for example
\cite{klempt,petry,1,3,Bugg,5} and references therein. The
question is whether $f_0(980)$ belongs to the scalar quark nonet
$1^3P_0 q\bar q$ or whether it should be considered as an exotic state.
Recent measurments of the radiative decay $\phi (1020) \to \gamma
f_0(980)$ \cite{6,7} have reinforced the discussion.

As was stressed in \cite{3,8,aanis}, it is reasonable to perform the
$q\bar q$ nonet classification in terms of the $K$-matrix, or "bare",
poles corresponding to states "before" the mixing which is caused by
the transitions $q\bar q\; state \to real\; mesons$. Such a mixing
is crucial for the formation of the scalar/isoscalar resonances in the
region 1200-1600 MeV, which are $f_0(1300)$, $f_0(1500)$ and the broad
state $f_0(1530^{+90}_{-250})$. According to the $K$-matrix analysis,
these resonances are formed as a result of mixing of the lightest
scalar gluonium with the $1^3P_0 q\bar q$ and $2^3P_0 q\bar q$ states.

In terms of $f_0^{bare}$, the scalar/isoscalar states of the basic
$1^3P_0q\bar q$ nonet are $f_0^{bare}(720\pm 100)$ and
$f_0^{bare}(1260\pm 30)$; the state $f_0^{bare}(720 )$ is close to
the flavour octet $\varphi=-69^\circ \pm 12^\circ$ (the mixing angle
is determined as $f_0=n\bar n \cos\varphi+s\bar s \sin\varphi$ where
$n\bar n =(u\bar u +d\bar d)/\sqrt 2$),
while the state $f_0^{bare}(1260)$ is almost a singlet
($\varphi=21^\circ\pm 12^\circ$). The transitions which are due to the
decay processes $f_0\to \pi\pi, K\bar K, \eta\eta$ mix the states
$f_0^{bare}(720)$ and $f_0^{bare}(1260)$ with each other as well as
with nearby states: $f_0^{bare}(1235\pm 50)$, $f_0^{bare}(1600\pm
50)$, $f_0^{bare}(1810\pm 30)$ (the orthoganality of the coordinate
wave functions does not work in the transition $(q\bar q-state)_1 \to
real\; mesons \to (q\bar q-state)_2$). Two of these states are
the $2^3P_0q\bar q$ nonet members and one is the scalar
gluonium, either $f_0^{bare}(1235 )$ or $f_0^{bare}(1600)$:
the $K$-matrix analysis cannot definitely tell us which one. Lattice
calculations \cite{lattice} favour $f_0^{bare}(1600)$ as a candidate
for the glueball. After the mixing, we have in the wave
$(IJ^{PC}=00^{++})$ four comparatively narrow resonances ($f_0(980),
f_0(1300),f_0(1500),f_0(1750)$) and one rather broad state
$f_0(1530^{+90}_{-250})$. The broad state $(\Gamma/2\simeq 400-500$
MeV) appears as a result of a specific phenomenon which is
appropriate for scalar/isoscalar states in the region 1200--1600 MeV,
that is the accumulation of widths of overlapping resonances by one of
them \cite{accum}. The analogous phenomenon in nuclear physics was
discussed in \cite{okun}. Fitting to data proves that
the main participants of the mixing are the glueball and two $q\bar q$
states belonging to the nonets $1^3P_0q\bar q$ and $2^3P_0q\bar q$,
whose flavour wave functions are close to singlets: the transition
$glueball \to q\bar q (singlet)$ is allowed,
while the transition
$glueball \to q\bar q (octet)$ is suppressed.
Gluonium/$q\bar q$ mixing is not suppressed according to the $1/N$
expansion rules \cite{t'Hooft} (see \cite{3,Anis_zeit} for
details). The glueball descendant is a broad state
$f_0(1530^{+90}_{-250})$ which contains about 40--50\% of the gluonium
component; the rest is shared mainly by $f_0(1300)$ and $f_0(1500)$.

Following the $K$-matrix scenario, the resonance $f_0(980)$ is the
descendant of $f_0^{bare}(720\pm 100)$: the shift of the bare pole
to the region of 1000 MeV is due to transitions of $f_0^{bare}(720\pm
100)$ into real two-meson states, $\pi\pi$ and $K\bar K$.
As to its origin, the state $f_0(980)$ is the
superposition of states $q\bar q$, $qq\bar q \bar q$, $K\bar K$ and
$\pi\pi$, and the wave function is to be represented by the
correspondent
Fock column. Concerning the $K\bar K$ and $\pi\pi$ components, one
should take into account only the part of wave functions which respond
to large meson-meson separations, $r_{meson}>R_{confinement}$;
on the contrary, for the $qq\bar q \bar q$ state
small interquark separations are to be taken into account.
As to the glueball admixture in $f_0(980)$, the relative
suppression of the decay $J/\Psi \to \gamma f_0(980)$ \cite{PDG-00}
tells us that it is not large.

To enlighten the problem of the content of $f_0(980)$, it would be
reasonable to suggest a dominance of certain component: basing
on the results of the $K$-matrix analysis \cite{3,8,aanis}
it is natural to check
whether the hypothesis about the dominance of $q\bar q$ component can
explain the data on radiative decays $\phi(1020) \to \gamma f_0(980)$
and $f_0(980) \to \gamma\gamma$. This study is dictated by the existing
opinion  that measured widths of radiative decays contradict the
interpretation of $f_0(980)$ as $q\bar q$ state, see mini-review
\cite{torn} and references therein.

In Section 2 we calculate the decay amplitude
$\phi(1020) \to
\gamma f_0(980)$ assuming $f_0(980)$ to be a $q\bar q$ system. The
technique for the description of composite $q\bar q$ systems and
corresponding form factor calculations has been presented in detail in
Ref. \cite{AMN}, where the pion form factor was studied as well
as transition form factors $\pi^0 \to \gamma \gamma^*(Q^2)$, $\eta \to
\gamma \gamma^*(Q^2)$ and $\eta' \to \gamma \gamma^*(Q^2)$.  The method
of spectral integration has been used for the form factor calculations.
Within this technique, it is possible to introduce normalized wave
functions. The transition amplitudes obey the
requirement of analyticity, causality and gauge invariance. The used
techique allows us also to perform calculations in terms of the
light-cone variables.

It is worth noting that this calculation technique for the processes
involving bound states has a broader applicability than for $q\bar q$
systems only: in \cite{AKMS} this method got its approbation by
describing the deuteron as a composite $np$ system, then this very
technique was applied to heavy mesons \cite{melikhov,melikhov-s}.
For the convenience of a reader, in Section 2 we recall briefly the
basic points of this approach and give necessary formulae for the
calculation of the reaction $\phi(1020) \to \gamma f_0(980)$.
Our calculations show that the data on branching ratio
$BR(\phi\to\gamma f_0(980))
=(3.4 \pm 0.4 ^{+1.5}_{- 0.5})\cdot 10^{-4}$ \cite{6,7,PDG-00}
may be described within the $q\bar q$ structure of $f_0(980)$,
with the mixing
of $s\bar s $ and
$n\bar n$ components as follows:
\begin{equation}
\psi_{flavour}(f_0(980))= n\bar n \cos\varphi+s\bar s \sin\varphi
\; , \qquad 25^\circ \le |\varphi|\le 90^\circ \; .
\end{equation}
The width $\Gamma (\phi\to\gamma f_0(980))$ depends strongly
on the radius squared of $f_0(980)$,  $R^2_{f_0(980)}$.
At large $R^2_{f_0(980)} \geq 12$ GeV$^{-2}$ ($0.47$ fm$^2$),
the data require $|\varphi|\sim 25^\circ -\; 45^\circ $, while at
$R^2_{f_0(980)} \sim 8$ GeV$^{-2}$ ($ 0.32$ fm$^2$) one needs
$|\varphi|\sim 40^\circ -\; 75^\circ $.

The estimation of radii of the scalar/isoscalar mesons, which has been
carried out by using GAMS data for $\pi^- p\to \pi^0\pi^0n$
\cite{GAMS}, shows that $q\bar q$ component in $f_0(980)$ is
comparatively compact, $R^2_{f_0(980)} = 6\pm 6$ GeV$^{-2}$
\cite{radii}. So,
basing on GAMS data, it is reasonable to put
\begin{equation}
6 \;{\rm GeV}^{-2}\le R^2_{f_0(980)} \le 12 \;{\rm GeV}^{-2} \ ,
\end{equation}
that favours the large values for the mixing angle.

Concerning the scheme under investigation, it is of principal
importance to verify if the transition $\phi(1020) \to \gamma
f_0(980)\to \gamma \pi^0\pi^0$ describes the measured $\pi^0\pi^0$
spectrum within the Flatt\'e formula \cite{Flatte}, with parameters
determined by hadronic reactions. In Section 3 we calculate the
$\pi^0\pi^0$ spectum using the parametrization of the Flatt\'e formula
suggested in \cite{Bugg,Andrei}, and reasonable agreement with data is
obtained. The performed spectrum calculations indicate the existence
of significant systematic errors in extracting the resonance signal
$\phi (1020)\to\gamma f_0(980)$ which are caused by the background
contribution.

Radiative decay $f_0(980)\to \gamma\gamma$ provides us another source
of information about the content of $f_0(980)$:
the calculation of partial width
$f_0(980)\to \gamma\gamma$ has been performed in \cite{f-gg} assuming
the $q\bar q$ structure of $f_0(980)$.
However, as was stressed in \cite{AA},
the extraction of the signal $f_0(980)\to\gamma\gamma$
from the measured spectra $\gamma\gamma\to\pi\pi$
meets with a difficulty, that is, a strong interference of the
resonance with background, thus giving uncontrollable errors. The
recently obtained partial width $\Gamma(f_0(980)\to \gamma\gamma) =
0.28 ^{+ 0.09}_{-0.13}$ keV \cite{Pennington} is a
factor two less than
the averaged value reported previously
($0.56 \pm 0.11$ keV \cite{PDG-98}). In Section 4 we re-analyse
the decay $f_0(980)\to \gamma\gamma$ using
the Boglione-Pennington number for the width
($0.28 ^{+0.09}_{-0.13}$ keV) together with the restriction for
the mean radius squared of $f_0(980)$ \cite{radii}:
$R^2_{f_0(980}) \le 12$ GeV$^{-2}$. This
provides two possible intervals for the $n\bar
n/s\bar s$ mixing angle:
\begin{equation}
f_0(980)\to \gamma\gamma: \qquad
1)\; 80^\circ \le \varphi \le
 93^\circ \; , \quad 2)\; (-54^\circ)\le \varphi \le (-42^\circ) \ .
\end{equation}

In Section 5 we calculate partial widths for
decays $\phi(1020) \to \gamma \eta,\gamma \eta',\gamma \pi^0$
in the framework of the same technique as is used for
$\phi(1020) \to \gamma f_0(980)$ and with the same
parametrization of the $\phi$-meson wave function. The calculations
demonstrate a good agreement with the data for these processes
as well.
 The calculation of the decays $\phi(1020) \to \gamma \eta,\gamma \eta'$
provides us with a strong argument that the applied method for
calculating radiative decays of $q\bar q$ mesons is wholly reliable,
and the decay $\phi(1020) \to \gamma \pi^0$ allows us to estimate the
 admixture of the
$n\bar n$ component in $\phi (1020)$. The partial width of
$\phi(1020) \to \gamma a_0(980)$ is also proportional to the
probability of the $n\bar n$ component in $\phi (1020)$, we discuss
this decay in Section 5 as well.

The restrictions for the mixing angle $\varphi$ which come from
the combined analysis of radiative decays
$\phi(1020)\to\gamma f_0(980)$ and $f_0(980)\to\gamma\gamma$ are
discussed in Section 6. For negative mixing angles the combined
analysis does not change the restriction (3):
\begin{equation}
\varphi=-48^\circ \pm 6^\circ \ ,
\end{equation}
and does not provide any restriction for
the radius $R^2_{f_0(980)}(n\bar n)$.

For positive mixing angles the allowed region of
$(\varphi, R^2_{f_0(980)}(n\bar n))$ reads as follows:
\begin{equation}
\varphi=86^\circ \pm 3^\circ \ , \qquad R^2_{f_0(980)}(n\bar n) \le
6.8\; {\rm GeV}^{-2} \ .
\end{equation}
This means that in this solution $f_0(980)$ should be rather compact
$s\bar s$ state.

The hadronic decays of $f_0(980)$ strongly contradict the almost
complete absence of the $n\bar n$ component in $f_0(980)$, that gives
the priority to the solution (4), according to which $f_0(980)$ is
close to the octet state (recall, $\varphi_{octet}=-54.7^\circ$).

\section{ $\phi (1020) \to \gamma \lowercase{f}_0(980) $:
the decay amplitude and partial width}

In this Section we calculate the coupling constant for the decay $\phi
\to \gamma f_0(980)$ assuming the $q\bar q$ structure of $ f_0(980)$.
In the calculation we use the spectral integration technique developed
in \cite{AMN}. Within the framework of this approach, we study the
reaction with a virtual photon, $\phi \to \gamma (q^2) f_0(980)$, and
then the limit $q^2 \to 0$ is attained.

\subsection{ Transition amplitude}

The transition amplitude $\phi \to \gamma (q^2) f_0(980)$ contains $S$
and $D$ waves. Correspondingly, the spin-dependent part of the
amplitude, $A_{\mu\nu}$ (the index $\mu$ stands for the spin of the
photon and $\nu$ to that of the $\phi$-meson), consists of two terms
which are proportional to $g_{\mu\nu}^\perp$ ($S$ wave) and $q^\perp
_\mu q^\perp _\nu -q^2_{\perp } g_{\mu\nu}^\perp /3$ ($D$ wave):
\begin{equation}
A_{\mu\nu}=e\left [g_{\mu\nu}^\perp F_S(q^2)+(q_\mu^\perp q_\nu^\perp-
\frac13 g_{\mu\nu}^\perp q^2_{\perp } )F_D(q^2) \right ].
\label{1.1}
\end{equation}
Here $q^\perp$ is the relative momentum of the produced particles (it
is orthogonal to the $\phi$ meson momentum $p$) and $g_{\mu\nu}^\perp$
is a metric tensor in the space orthogonal to $p$:
\begin{equation}
q_\mu^\perp=q_\mu-p_\mu \frac{(qp)}{p^2}\ , \qquad
g_{\mu\nu}^\perp=g_{\mu\nu}-\frac{p_\mu p_\nu}{p^2} \ .
\label{1.2}
\end{equation}
The requirement $q_\mu A_{\mu\nu}=0$
results in the following constraint for the
invariant amplitudes $F_S(q^2)$ and $F_D(q^2)$:
\begin{equation}
F_S(q^2)+\frac23 (qq^\perp)F_D(q^2)=0 \ ,
\label{1.3}
\end{equation}
hence equation (\ref{1.1}) can be re-written as
\begin{equation}
A_{\mu\nu}=\frac 32 e F_S(q^2) \left (g_{\mu\nu}^\perp -
\frac{ q_\mu^\perp q_\nu^\perp}{q^2_{\perp} } \right )
\equiv e A_{\phi \to \gamma f_0}(q^2)
\; g^{\perp\perp}_{\mu\nu} \ .
\label{1.4}
\end{equation}
The metric tensor $g^{\perp\perp}_{\mu\nu}$ works in the space
orthogonal to the reaction momenta: $g^{\perp\perp}_{\mu\nu} p_\nu =0$
and $g^{\perp\perp}_{\mu\nu} q_\mu =0$.

\subsection{ Partial width}

The decay partial width is determined as
\begin{equation}
m_\phi \Gamma_{\phi \to f_0\gamma} =\frac13 \int
d\Phi(p;q,k_f) |A_{\mu\nu}|^2\ .
\label{1.5}
\end{equation}
In (\ref{1.5}) the averaging over spin projections of the
$\phi$-meson and summing over photon ones are carried out; $d\Phi$ is
the two-particle phase space which is defined as follows:
\begin{equation}
d\Phi(p;k_1,k_2) = \frac12 \frac{d^3k_1}{(2\pi)^3 2k_{10}}
\frac{d^3k_2}{(2\pi)^3 2k_{20}} (2\pi)^4 \delta^{(4)}(p-k_1-k_2) \ .
\label{1.6}
\end{equation}
For the radiative decay $\phi \to \gamma+f_0(980)$, one has $\int
d\Phi(p;q,k_f)=(m_\phi ^2-M_f^2)/(16\pi m_\phi ^2)$.

After summing over spin variables,
the partial width reads:
\begin{equation}
m_\phi \Gamma_{\phi \to f_0\gamma}=\frac16 \alpha
\frac{m_\phi ^2-M_f^2}{m_\phi ^2} |A_{\phi \to \gamma f_0} (0)|^2 .
\label{1.9}
\end{equation}
Here $\alpha=e^2/4\pi=1/137$.

\subsection{Double spectral representation
for the transition form factor $\phi(1020)\to\gamma f_0(980)$}

Assuming the $q\bar q$ structure of $f_0(980)$, the coupling constant
of the decay $\phi \to \gamma (q^2) f_0(980)$ is determined by the
processes $\phi \to q\bar q$
and $q\bar q \to f_0(980)$, with the emission
of a photon by the quark in the intermediate state, see Fig. 1a.
Following the prescriptions of Ref. \cite{AMN}, we calculate the
three-point quark loop shown in Fig. 1b by using the double spectral
representation over intermediate $q\bar q$ states marked by dashed
lines in Fig. 1b.

To be illustrative, let us start with the three-point
Feynman diagram.

\subsubsection{Double spectral integral for three-point quark
diagram}

For the process $V\to \gamma S$,
where $V$ and $S$ stand for scalar and vector particle,
the Feynman diagram of Fig. 1a reads:
\begin{equation}
A_{\mu\nu}^{(Feynman)}
=\int \frac{d^4k}{i(2\pi)^4}\; G_V \;
\frac{Z^{(q\bar q)}_{V\to \gamma S} \; S^{(V\to S)}_{\mu\nu}}
{(m^2-k_1^2)(m^2-k'^2_1) (m^2-k^2_2)}\; G_S\ .
\label{2.1}
\end{equation}
Here
$k_1$, $k'_1$, $k_2$ are quark momenta,
$m$ is the quark mass, and $G_V$, $G_S$ are
quark-meson vertices; the quark charges are included into
 $Z^{(q\bar q)}_{V\to \gamma S}$.
The spin-dependent factor reads:
\begin{equation}
S^{(V \to S)}_{\mu\nu}=-{\rm Sp} \left [
(\hat k'_1+m)\gamma^\perp_\mu (\hat k_1+m)\gamma^\perp_\nu
(-\hat k_2+m) \right ] \ ,
\label{2.2}
\end{equation}
where the Dirac matrices
$\gamma^\perp_\mu $ and  $\gamma^\perp_\nu $ are orthogonal to the
emmitted momenta:
$\gamma^\perp_\mu q_\mu =0 $ and $\gamma^\perp_\nu p_\nu =0$.

To transform the Feynman integral
(\ref{2.1}) into the double spectral intergral over invariant
$q\bar q$ masses squared,
one should make the following steps:\\
(i)  consider the corresponding energy-off-shell diagram, Fig. 1b,
with $P^2=(k_1+k_2)^2 \ge 4m^2$, $P'^2=(k'_1+k_2)^2 \ge 4m^2$ and
fixed momentum transfer squared $q^2 =(P-P')^2$,\\
(ii) extract the invariant amplitude separating spin operators,\\
(iii) calculate the discontinuities of the invariant amplitude
over intermediate $q\bar q$ states marked in Fig. 1b by dashed lines.

The double discontinuity is the integrand of the spectral
integral over $P^2 $ and $P'^2 $.
Furthermore, we put the following notations:
\begin{equation}
P^2=s, \qquad P'^2=s' \ .
\label{2.3}
\end{equation}
For the calculation of discontinuity, by cutting Feynman diagram,
the pole terms of the propagators are replaced with
their residues:  $(m^2-k^2)^{-1} \to \delta (m^2-k^2)$. So, the
particles in the intermediate states
marked by dashed lines I and II in Fig. 1b
are mass-on-shell,
$k^2_1=k^2_2=k'^2_1=m^2$. As a result, the Feynman diagram
integration turns into the integration over phase spaces of the cut
diagram states.
Corresponding phase space for the three-point diagram reads:
\begin{equation}
d\Phi(P,P';k_1,k_2,k'_1) = d\Phi(P;k_1,k_2)d\Phi(P';k'_1,k'_2)
(2\pi)^3 2k_{20}\delta^{(3)}(\vec k'_2-\vec k_2) \ ,
\label{2.4}
\end{equation}
where the invariant two-particle phase space
$d\Phi(P;k_1,k_2)$ is determined as follows:
\begin{equation}
d\Phi(P;k_1,k_2) = \frac12 \frac{d^3k_1}{(2\pi)^3 2k_{10}}
\frac{d^3k_2}{(2\pi)^3 2k_{20}} (2\pi)^4 \delta^{(4)}(P-k_1-k_2) \ .
\label{2.5}
\end{equation}
The last step is to single out the invariant component
from the spin factor (\ref{2.2}).
According to (\ref{1.4}), the spin factor (\ref{2.2}) is proportional
to the metric tensor:
$S^{(V \to S)}_{\mu\nu} \sim g^{\perp\perp}_{\mu\nu} $.
Then the spin factor $S^{(tr)}_{V\to\gamma S}$, determined as
\begin{equation}
S^{(V \to S)}_{\mu\nu} =
g^{\perp\perp}_{\mu\nu}S^{(tr)}_{V\to\gamma S}(s,s', q^2),
\label{2.6}
\end{equation}
is equal to:
$$
S^{(tr)}_{V\to\gamma S}(s,s', q^2)=
-2m \left (4m^2 +s-s'-\frac{4ss'}{s+s'-q^2}\alpha(s,s',q^2 \right ),
$$
\begin{equation}
\alpha(s,s',q^2)= \frac{q^2(s+s'-q^2)}{2q^2(s+s')-(s-s')^2-q^4}.
\label{2.7s}
\end{equation}
Here we have used that
$(k_1k_2)=s/2-m^2$, $(k'_1k_2)=s'/2-m^2$ and $(k'_1k_1)=m^2-q^2/2$.

The spectral integration is carried out over the energy squared of
quarks in the intermediate states,
$s=P^2=(k_1+k_2)^2$ and $s'=P'^2=(k'_1+k_2)^2$, at fixed $q^2=(P'-P)^2$.
The spectral
representation for the amplitude
$A_{\mu\nu}^{(Feynman)}=g^{\perp\perp}_{\mu\nu}
A_{V\to \gamma S}^{(Feynman)}(q^2)$ reads:
\begin{eqnarray}
A^{(Feynman)}_{V\to \gamma S}(q^2)
&=&\int \limits_{4m^2}^\infty \frac{ds}{\pi} \int
\limits_{4m^2}^\infty \frac{ds'}{\pi}
\frac{G_V}{s-M_V^2}\frac{G_S}{s'-M_S^2}
\nonumber \\
&\times& \int d\Phi(P,P';k_1,k_1',k_2)\,
S^{(tr)}_{V\to \gamma S}(s,s', q^2)
Z^{(q\bar q)}_{V\to\gamma S}\ .
\label{2.7}
\end{eqnarray}
It is reasonable to name the ratios $G_V/(s-M^2_V)$ and
$G_S/(s'-M^2_S)$ the wave functions of vector and scalar particle,
respectively:
\begin{equation}
\frac{G_V}{s-M^2_V}=\psi_V(s), \qquad
\frac{G_S}{s'-M_S^2}=\psi_S(s')\ .
\label{2.8}
\end{equation}
Equation (\ref{2.7}) is a spectral representation of the Feynman
amplitude shown in Fig. 1a for point-like vertices. In general case it
is necessary to take account of the $s$-dependent vertices of the
reaction.

\subsubsection{Form factor $\phi(1020)\to\gamma f_0(980)$}

Generally, the energy-dependent vertices
can be incorporated into spectral integrals. According to
\cite{AMN,AKMS}, the form factor of a composite system can be obtained
by considering the two-particle partial-wave  scattering amplitude
$1+2\to 1+2$, with the same quantum numbers as for the composite system.
In this amplitude, the composite system reveals itself as a pole.
The amplitude for the emission of a photon by the
two-particle-interaction system has two poles related to the states
{\it before} and {\it after}
electromagnetic interaction, and the two-pole residue of this amplitude
provides us form factor of the composite system. When a
partial-wave scattering amplitude is treated using the
dispersion relation $N/D$-method, the vertex $G(s)$ is determined by
the $N$-function:  the vertex as well as $N$-function have
left-hand-side singularities which are due to forces between particles
$1+2$.

Generally, the form factor for the tansition $\phi(1020) \to
\gamma(q^2) f_0(980)$ reads:
\begin{eqnarray}
A_{\phi\to \gamma f_0}(q^2)
&=&\int \limits_{4m^2}^\infty \frac{ds}{\pi} \int
\limits_{4m^2}^\infty \frac{ds'}{\pi}
\psi_\phi (s)\psi_{f_0} (s')
\nonumber \\
&\times& \int d\Phi(P,P';k_1,k_1',k_2)\,
S^{(tr)}_{\phi\to \gamma f_0}(s,s', q^2)
Z^{(q\bar q)}_{\phi\to\gamma f_0}\ .
\label{2.7a}
\end{eqnarray}
The factor $Z^{(q\bar q)}_{\phi\to f_0}$ is defined by
the flavor meson component: for $q\bar q=n\bar n$ we have $Z^{(n\bar
n)}_{\phi\to f_0}=\sqrt{2}/3$ and for the $s\bar s$ component
$Z^{(s\bar s)}_{\phi\to f_0}=-2/3$.

Working with Eqs. (\ref{2.7a}), one can either\\
(i) express it in terms of the light cone variables, or\\
(ii) keep the spectral integrals over $s$ and $s'$ and eliminate
integrations over quark momenta with the help of the phase space
$\delta$-functions.

{\bf a) Light-cone variables}

The  formulae (\ref{2.4}), (\ref{2.5}), (\ref{2.7a}) allow one
to make the  transformation to
the light-cone variables, using the boost along the $z$-axis.
Let us use the frame
where the initial vector meson  moves along the
$z$-axis with the momentum
$p\to \infty$:
\begin{equation}
P=(p+\frac {s}{2p}, 0, p), \qquad
P\; '=(p+\frac {s'+q^2_{\perp}}{2p}, -\vec q_{\perp}, p).
\label{2.9}
\end{equation}
In this frame the two-particle phase space is equal to
\begin{eqnarray}
d\Phi(P;k_1,k_2)&=&\frac{1}{16\pi^2} \frac{dx_1dx_2}{x_1x_2}
d^2k_{1\perp}d^2k_{2\perp} \delta (1-x_1-x_2 )
\delta^{(2)}(\vec k_{1\perp}+\vec k_{2\perp})
\nonumber\\
&\times&\delta \left (s-\frac{m^2+k^2_{1\perp}}{x_1} -
\frac{m^2+k^2_{2\perp}}{x_2} \right )\ ,
\label{2.10}
\end{eqnarray}
and the phase space for the triangle diagram reads:
\begin{eqnarray}
d\Phi(P,P';k_1,k_2,k'_1)&=&
\frac{1}{16\pi} \frac{dx_1dx_2}{x^2_1x_2}
d^2k_{1\perp}d^2k_{2\perp} \delta (1-x_1-x_2 )
\delta^{(2)}(\vec k_{1\perp}+\vec k_{2\perp})
\nonumber\\
&\times&
\delta \left (s-\frac{m^2+k^2_{1\perp}}{x_1}
-\frac{m^2+k^2_{2\perp}}{x_2} \right )
\nonumber\\
&\times&
\delta \left (s'+q^2_\perp-\frac{m^2+(\vec k_{1\perp}-\vec q_{\perp})
^2}{x_1}
-\frac{m^2+k^2_{2\perp}}{x_2} \right )\ .
\label{2.11}
\end{eqnarray}
After having integrated over $s$ and $s'$, by using $\delta$-functions,
we have:
\begin{equation}
A_{\phi \to \gamma(q^2) f_0  } (q^2)=
\frac {Z_{\phi \to \gamma f_0 }^{(q\bar q)} }{16\pi^3}
\int \limits_{0}^{1}
\frac {dx}{x(1-x)^2}  \int d^2k_{\perp} \psi_{\phi} (s)
\psi_{f_0}(s') S^{(tr)}_{\phi \to \gamma f_0 } (s,s',q^2)\ ,
\label{2.12}
\end{equation}
where $ x=k_{2z}/p$ , $ \vec k_{\perp}=  \vec k_{2\perp}$, and
the $q\bar q$ invariant energies squared are
\begin{equation}
s=\frac{m^2+k^2_{\perp} }{x(1-x)}, \qquad
s'=\frac{m^2+(\vec k_{\perp}+x \vec q_{\perp})^2 }{x(1-x)}\ .
\label{2.13}
\end{equation}

{\bf b) Spectral integral representation}

In Eq. (\ref{2.7a}) one
can integrate over the phase space keeping fixed the
energies squared, $s$ and $s'$. After using the
phase space $\delta$-functions, we have:
\begin{eqnarray}
A_{\phi \to \gamma f_0 }(q^2)&=&\int
\limits_{4m^2}^\infty \frac{dsds'}{\pi^2}
\psi_\phi (s)\psi_{f_0}(s') \frac
{\theta\left (ss'Q^2-m^2\lambda(s,s',Q^2)\right )}
{16\sqrt{\lambda(s,s',Q^2)}}
\nonumber\\
&\times&
Z^{(q\bar q)}_{V\to\gamma S}S^{(tr)}_{V \to\gamma S}(s,s',Q^2),
\label{2.14}
\end{eqnarray}
with
\begin{equation}
\lambda(s,s',Q^2)= (s'-s)^2+2Q^2(s'+s)+Q^4\ .
\label{2.15}
\end{equation}
The $\theta$-function, $\theta (X)$, restricts the integration region
for different $Q^2=-q^2$: $\theta(X)=1$ at $X\ge 0$ and $\theta(X)=0$
at $X < 0 $.

In the limit $Q^2 \to 0$, the integration over $s'$ is carried
out, and we have:
\begin{equation}
A_{\phi \to \gamma f_0 }(0)=\int
\limits_{4m^2}^\infty \frac{ds}{\pi} \psi_\phi (s)\psi_{f_0}(s)
\left [ -\frac{m^3}{2\pi} \ln
\frac{\sqrt{s}+\sqrt{s-4m^2}}{\sqrt{s}-\sqrt{s-4m^2}}
+\frac{m}{4\pi}\sqrt{s(s-4m^2)} \right ]
Z^{(q\bar q)}_{\phi \to \gamma f_0 }\ .
\label{2.16}
\end{equation}
To calculate
$A_{\phi \to \gamma f_0} (0)$
one should determine the wave functions of the
$\phi$-meson and $f_0(980)$.

\subsubsection{ Wave functions of $\phi(1020)$ and $f_0(980)$.}

We write down the wave functions of $\phi(1020)$ and $f_0(980)$
expanding them in a series as follows:
\begin{eqnarray}
&&\Psi_\phi(s)=\left (n\bar n\sin \varphi_V + s\bar s\cos \varphi_V
\right ) \psi_\phi(s)\ , \nonumber \\
&&\Psi_f(s)=\left ( n\bar n\cos \varphi+ s\bar s
\sin \varphi\right ) \psi_f(s) \ .
\end{eqnarray}
For $\psi_\phi(s)$ and $\psi_f(s)$ the exponential parametrization is
used:
\begin{equation}
\psi_\phi(s)=C_\phi e^{-b_\phi s}, \qquad
\psi_f(s)=C_f e^{-b_f s}\ .
\label{exp}
\end{equation}
The parameters $b_\phi$ and $b_f$ characterise the size of the system,
they are related to the mean radii squared of these mesons,
$R^2_\phi$ and $R^2_{f_0(980)}$ and $C_\phi $ and $C_f$
are the normalization constants. In our approach the dynamics of
interaction is hidden in the wave functions  $\psi_{\phi}$ and
$\psi_{f_0}$. We assume exponential $s$-dependence for the wave
functions. Recall that exponential parametrization is often used in the
quark model. Working upon meson radiative decays, we have additional
arguments to employ this presentation of wave functions of the
low-lying $q\bar q$ states.

In Ref. \cite{AMN}, in our analysis of pion form factor as well as
transition form factors  $\pi^0 \to \gamma \gamma^*(Q^2)$, $\eta \to
\gamma \gamma^*(Q^2)$ and $\eta' \to \gamma \gamma^*(Q^2)$, we have
reconstructed wave functions of pseudoscalar mesons as functions of
$s$. At small relative momenta of quark and antiquark, $k^2 \le 0.6$
(GeV/c)$^2$ (where $k^2=s/4-m^2$), the wave functions demonstrate rapid
-- exponential-type -- decrease, then it becomes slower. The same
behaviour is appropriate to wave function of the photon, which was used
in our calculations; it is shown below, in Fig. 5). We keep in mind
just this form of the wave functions for low-lying $q\bar q$ mesons,
that is a foundation for the formula (\ref{exp}).
In \cite{eta}, at
calculation of the transitions $\eta \to \gamma\gamma$, $\eta' \to
\gamma\gamma$, $\eta(1295) \to \gamma\gamma$ and $\eta(1440) \to
\gamma\gamma$ we checked the possibility to use exponential-type wave
functions. The processes $\eta \to \gamma\gamma$ and $\eta' \to
\gamma\gamma$ have been calculated both with the wave function
reconstructed from the experiment and exponential-type wave
functions, with the same $R^2$. The difference between form factors is
not greater than 5\% .

At fixed $R^2_\phi$ and $R^2_{f_0(980)}$ the constants $C_\phi$ and
$C_f$ are determined by the wave function normalization, which itself
is given by the meson form factor in the external field,
$F_{meson}(q^2)$, namely, at small $q^2$:
\begin{equation}
F_{meson}(q^2)\simeq 1+\frac16 R^2_{meson}q^2 \ .
\end{equation}
The requirement $F_{meson}(0)=1$ fixes the constant $C_{meson}$
in (\ref{exp}), while the value $R^2_{meson}$ is directly related
to the parameter $b_{meson}$.

The form factor $F_{meson}(q^2)$ reads \cite{AMN}:
\begin{equation}
F_{meson}(q^2)=\int\limits_{4m^2}^\infty \frac{dsds'}{\pi^2}
\psi_{meson}(s)\psi_{meson}(s')\frac
{\theta\left (ss'Q^2-m^2\lambda(s,s',Q^2)\right )}
{16\sqrt{\lambda(s,s',Q^2)}}S^{(tr)}_{meson}(s,s',q^2) \ ,
\end{equation}
where $S^{(tr)}_{meson}$ for $f_0(980)$ and $\phi(1020)$ is
determined by the following traces:
\begin{eqnarray}
&&2P^\perp_\mu S^{(tr)}_{f_0}(s,s',q^2)=-{\rm Sp}\left
[(\hat k'_1+m)\gamma^\perp_\mu (\hat k_1+m)(-\hat k_2+m)\right ]\ ,
\nonumber \\
&&2P^\perp_\mu S^{(tr)}_{\phi}(s,s',q^2)=- \frac13{\rm Sp}\left
[\gamma'^\perp_\alpha(\hat k'_1+m)\gamma^\perp_\mu (\hat k_1+m)
\gamma^\perp_\alpha(-\hat k_2+m)\right ] \ .
\end{eqnarray}
Here the orthogonal components are as follows:
\begin{eqnarray}
&&P^\perp_\mu =P_\mu -q_\mu\frac {(Pq)}{q^2}\ , \qquad
\gamma_\mu^\perp=\gamma_\mu-q_\mu \frac {\hat q}{q^2}\ ,
\nonumber \\
&&\gamma_\alpha^\perp=\gamma_\alpha-P_\alpha \frac {\hat P}{P^2}\ ,
\qquad
 \gamma'^\perp_\alpha=\gamma_\alpha-P'_\alpha \frac {\hat P'}{P'^2}\ ,
\label{23}
\end{eqnarray}
and $q=k'_1-k_1$.
When determining $S^{(tr)}_\phi$, we have averaged over
polarizations of the $\phi$ meson (the factor
$1/3$).

The functions $ S^{(tr)}_{f_0}$ and $S^{(tr)}_{\phi}$ are
equal to:
\begin{eqnarray}
&&S^{(tr)}_{f_0}(s,s',q^2)=\alpha(s,s',q^2)(s+s'-8m^2-q^2)+q^2\ ,
\nonumber \\
&&S^{(tr)}_{\phi}(s,s',q^2)=
\frac23 \left [\alpha(s,s',q^2)(s+s'+4m^2-q^2)+q^2
\right ]\ .
\label{24}
\end{eqnarray}

The radius squared of the $n\bar n$ component in the $\phi$-meson is
suggested to be approximately the same as that of the pion:
$R^2 _\phi (n\bar n) \simeq 10.9$ GeV$^{-2}$, while the radius
squared $R^2 _\phi (s\bar s)$ is slightly less, $R^2
_\phi (s\bar s) \simeq 9.3$ GeV$^{-2}$ (that corresponds to $b_\phi =
2.5$ GeV$^{-2}$). As to $f_0(980)$, we vary its radius in the interval
$6\;{\rm GeV}^{-2} \le R^2 _{f_0(980)} \le 18\;{\rm GeV}^{-2}$.

\subsubsection{ The result for $\phi(1020)\to\gamma +f_0(980)$.}

The amplitude $A_{\phi \to \gamma f_0} (0)$
is a sum of two terms related to the
$n\bar n$ and $s\bar s$ components:
\begin{equation}
A_{\phi \to \gamma f_0} (0)
=\cos \varphi\sin \varphi_V A^{(n\bar n)}_{\phi \to \gamma f_0} (0)+
\sin \varphi\cos \varphi_V A^{(s\bar s)}_{\phi \to \gamma f_0} (0)\ .
\label{25}
\end{equation}
In our estimations we put $\cos \varphi_V \sim 0.99$ and,
correspondingly, $|\sin \varphi_V| \le 0.1$;
 for $f_0(980)$ we vary the mixing angle in the interval
$0^\circ \le |\varphi|\le 90^\circ$.

The results of the calculation are shown in Figs. 2 and 3. In Fig. 2
the values $A^{(n\bar n)}_{\phi \to \gamma f_0} (0)$ and $A^{(s\bar
s)}_{\phi \to \gamma f_0} (0)$ are plotted with respect to the radius
squared $R^2_{f_0(980)}$, while the mean radius squared of the
$\phi$-meson is close to that of the pion: $R^2_{\phi(1020)} \simeq
0.4$ fm$^2$.

In Fig. 3 one can see the value $BR(\phi\to\gamma f_0(980))$ at
various $\sin|\varphi|$: $\sin|\varphi|$=0.4, 0.6, 0.8, 0.9, 0.985.
Shaded areas
correspond to the variation of $\varphi_V$ in the interval $-8^\circ
\le \varphi_V \le 8^\circ$; the lower and upper curves of the shaded
area correspond to destructive and constructive interferences of
$A^{(n\bar n)}_{\phi \to \gamma f_0} (0)$ and
$A^{(s\bar s)}_{\phi \to \gamma f_0} (0)$,
respectively.

The measurement of the $f_0(980)$ signal in the $\gamma \pi^0\pi^0$
reaction gives the branching ratio
$BR(\phi\to\gamma f_0(980))=(3.5\pm 0.3 ^{+1.3}_{-0.5})\cdot10^{-4}$
\cite{7}, in the analysis of $\gamma \pi^0\pi^0$ and
$\gamma \pi^+\pi^-$ channels it was found
$BR(\phi\to\gamma f_0(980))=
(2.90\pm 0.21 \pm 1.5)\cdot10^{-4}$ \cite{6}; the averaged
value is given in \cite{PDG-00}:
$BR(\phi\to\gamma f_0(980))=(3.4\pm 0.4)\cdot10^{-4}$.
Our calculations of the $\pi^0\pi^0$ spectrum with the
conventional Flatt\'e formula (Section 4) support the statement of
\cite{6,7} about the presence of large systematic errors
related to the procedure of  extracting
 the $f_0(980) $ signal. In our
estimation of the permissible interval for the mixing angle $\varphi$,
we have used the averaged value given by \cite{PDG-00}, with the
inclusion of systematic errors of the order of those found in
\cite{6,7}:  $BR(\phi\to\gamma f_0(980))=(3.4\pm 0.4
^{+1.5}_{-0.5})\cdot10^{-4}$ .

The calculated values of $BR(\phi\to\gamma f_0(980))$
agree with experimental data for
$|\varphi|\ge 25^\circ$; larger values of the mixing angle, $|\varphi|$
$\geq 55^\circ$,
correspond to a more compact structure of $f_0(980)$, namely,
$R^2_{f_0(980)} \leq 10$ GeV$^{-2}$, while small mixing
angles $|\varphi|\sim 25^\circ$ are related to a loosely bound
structure, $R^2_{f_0(980)} \geq 12$ GeV$^{-2}$ (recall that for the
pion $R^2_\pi \simeq 10$ GeV$^{-2}$).

The evaluation of the radius
of $f_0(980)$ was performed in \cite{f-gg} on the basis of GAMS data
\cite{GAMS}, where the $t$-dependence was measured in the process
$\pi^-p \to f_0(980)\,n$:
these data favour the comparatively compact structure of
the $q\bar q$ component in $f_0(980)$, namely,
$ R^2_{f_0(980)} =6\pm 6$ GeV$^{-2}$.

\section{Calculation of the partial width for
$\phi(1020) \to \gamma\pi^0\pi^0$}

In the $\pi^0\pi^0$ spectrum, the resonance $f_0(980)$ is seen as a
peak on the edge of the spectrum. So it is rather instructive to
calculate the dependence of $BR(\phi(1020) \to \gamma\pi^0\pi^0)$ on
$M_{\pi\pi}$ by using the conventional Flatt\'e formula \cite{Flatte}.

The resonance $ f_0(980)$ in hadronic reactions is well described by
the Flatt\'e formula which takes into account two decay
channels, $ f_0(980)\to\pi\pi$ and $f_0(980)\to K\bar K$;
the parameters of the Flatt\'e formula may be found in
\cite{Bugg,Andrei}.

In the calculation of the $\pi\pi$ spectrum, we take into account
the resonance contribution and the background, $B(M^2_{\pi\pi})$.
The partial width
$d\Gamma_{\phi\to\gamma \pi\pi}$ reads:
\begin{eqnarray}
d\Gamma_{\phi\to\gamma \pi\pi}&=&\Gamma_{\phi\to\gamma f_0(980)} \;
\frac{M_f(m^2_\phi-M^2_{\pi\pi})}{M_{\pi\pi}(m^2_\phi-M^2_f)}
\;\frac{F^2_{thresh}(M^2_{\pi\pi})}{F^2_{thresh}(M^2_f)}
\nonumber \\
&\times& \left | \frac{1 }{M^2_0-M^2_{\pi\pi}-ig^2_\pi\rho_{\pi\pi}-
ig^2_K\rho_{K\bar K} } + B(M^2_{\pi\pi}) \right |^2
g^2_\pi\rho_{\pi\pi} \frac{dM_{\pi\pi}^2}{\pi}\ ,
\label{26}
\end{eqnarray}
where $M_f$=0.98 GeV.
$F_{thresh}(M^2_{\pi\pi})$ is a factor which provides the threshold
behaviour of the decay amplitude $\phi(1020) \to \gamma\pi^0\pi^0$ at
$M^2_{\pi\pi}\to m^2_\phi$. We have chosen it as
$F^2_{thresh}(M^2_{\pi\pi})=1-\exp
[-(M^2_{\pi\pi}-m^2_\phi)^2/\mu^4_0]$, where $\mu_0$ is a scale
parameter; below we put $\mu_0=2m_\pi$.

Here we separate
$\Gamma_{\phi\to\gamma f_0(980)}$ and
introduce the ratio of
phase spaces for $(\pi\pi)+\gamma$ and $f_0(980)+\gamma$:
\begin{equation}
\frac{d\Phi_{(\pi\pi)\gamma}}{d\Phi_{f_0(980)\gamma}}=
\frac{M_f(m^2_\phi-M^2_{\pi\pi})}{M_{\pi\pi}(m^2_\phi-M^2_f)}\ .
\label{27}
\end{equation}
The values $g_\pi$ and $g_K$ are coupling constants for the
transitions $f_0(980)\to \pi\pi,K\bar K$. Also $\rho_{\pi\pi}$ and
$\rho_{K\bar K}$ are phase spaces for the $\pi\pi$ and $K\bar K$
states: $\rho_{\pi\pi}=\sqrt{M^2_{\pi\pi}-4m^2_\pi}/(16\pi
M_{\pi\pi})$ and $\rho_{K\bar K}=\sqrt{M^2_{\pi\pi}-4m^2_K}/(16\pi
M_{\pi\pi})$; recall that in the case of $M^2_{\pi\pi}< 4m^2_K$ the
kaon phase space is imaginary, for $\sqrt{ M^2_{\pi\pi}- 4m^2_K } \to
i\sqrt{4m^2_K-M^2_{\pi\pi} }$.

According to \cite{Bugg,Andrei}, the Flatt\'e formula is parametrized
as follows:
\begin{equation}
g^2_\pi\rho_{\pi\pi}\to (0.12\,{\rm GeV})\sqrt{M^2_{\pi\pi}-4m^2_\pi},
\qquad
g^2_K \rho_{K\bar K}\to (0.27\,{\rm GeV})\sqrt{M^2_{\pi\pi}-4m^2_K}\ ,
\label{28}
\end{equation}
$$
M_0 =0.99\pm 0.01\; {\rm GeV}.
$$
We parametrize the background term as
$B(M^2_{\pi\pi}) =a+bM^2_{\pi\pi}$ where $a$ and $b$ are
complex constants. The unitarity condition in the $\pi\pi$ channel
tells us that the decay amplitude $\phi(1020)\to\gamma\pi\pi$ has the
complex phase factor related to the $\pi\pi$ phase shift in the
$(IJ^{PC}=00^{++})$ channel:
$A_{\gamma\pi\pi}=|A_{\gamma\pi\pi}|\exp{[i\delta^0_0(M_{\pi\pi})]}$.
Therefore, one has
\begin{eqnarray}
&&\left | \frac{1 }{M^2_0-M^2_{\pi\pi}-ig^2_\pi\rho_{\pi\pi}-
ig^2_K\rho_{K\bar K} } + B(M^2_{\pi\pi}) \right |
e^{i\delta^0_0 (M_{\pi\pi})}
\nonumber \\
=&&\left ( \frac{1 }{M^2_0-M^2_{\pi\pi}-ig^2_\pi\rho_{\pi\pi}-
ig^2_K\rho_{K\bar K} } + B(M^2_{\pi\pi}) \right )
e^{i\Delta\delta^0_0 (M_{\pi\pi})}.
\label{5.4}
\end{eqnarray}
We parametrize the phase $\Delta\delta^0_0 (M_{\pi\pi})$ as
$\Delta\delta^0_0 (M_{\pi\pi})=\Delta_0+\Delta_1 (M_{\pi\pi}/m_0-1)
+\Delta_2 (M_{\pi\pi}/m_0-1)^2$ with $m_0$= 1 GeV.
The complexity of $B(M^2_{\pi\pi})$ is determined by the difference of
phases in the terms for the $f_0(980)$ production and  primary
background contribution.

To calculate the $\gamma\pi^0\pi^0$ spectrum, one should
multiply (\ref{26})
by the factor related to the $\pi^0\pi^0$ channel
$d\Gamma_{\phi\to\gamma \pi^0\pi^0}=\frac13 d\Gamma_{\phi\to\gamma
\pi\pi}$.
Figure 4 demonstrates $BR(\phi\to\gamma \pi^0\pi^0)$
calculated using Eq. (\ref{26}), with
$BR(\phi\to\gamma f_0(980))=3.4\cdot 10^{-4}$, $5.4\cdot 10^{-4}$
and $M_0 = 0.98$ GeV, $1.00$ GeV.
The following parameters (in GeV units) are used for the background
term $B(M^2_{\pi\pi})$ and the phase $\Delta\delta^0_0 (M_{\pi\pi})$:
\begin{eqnarray}
(a)\;{\rm BR}=3.4\times 10^{-4}\,,\; M_0=0.98:\quad &&a=-1.24+i0.74\;,
b=1.91-i0.95\, ,
\nonumber \\  &&
\Delta_0=      99^\circ\;,
\Delta_1=      11^\circ\;,
\Delta_2=     440^\circ\;;
\nonumber \\
(b)\;{\rm BR}=3.4\times 10^{-4}\,,\; M_0=1.00:\quad &&a=-1.96+i0.073\;,
b=2.37+i1.81\, ,
\nonumber \\  &&
\Delta_0=      106^\circ\;,
\Delta_1=     -136^\circ\;,
\Delta_2=      341^\circ\;;
\nonumber \\
(c)\;{\rm BR}=5.4\times 10^{-4}\,,\; M_0=0.98:\quad &&a=-0.84+i0.91\;,
b=0.73-i2.63\, ,
\nonumber \\  &&
\Delta_0=       98^\circ\;,
\Delta_1=      203^\circ\;,
\Delta_2=      735^\circ\;;
\nonumber \\
(d)\;{\rm BR}=5.4\times 10^{-4}\,,\; M_0=1.00:\quad &&a=-1.13+i0.69\;,
b=1.39-i1.13\, ,
\nonumber \\  &&
\Delta_0=       94^\circ\;,
\Delta_1=        9^\circ\;,
\Delta_2=      462^\circ\;.
 \label{30}
\end{eqnarray}
It is seen that the data \cite{7} agree reasonably with the Flatt\'e
parametrization for hadronic reactions given by \cite{Bugg,Andrei}. A
good description of the $\pi^0\pi^0$ spectrum at $BR(\phi\to\gamma
f_0(980))=5.4\cdot 10^{-4}$ supports the statement of \cite{6,7} about
large systematic errors in the determination of the $\gamma f_0(980)$
signal.

\section{Radiative decay $\lowercase{f}_0(980)\to \gamma \gamma $}

The amplitude for the transition $f_0(980)\to \gamma (q^2) \gamma
(q'^2)$ has the same spin structure as $\phi \to \gamma f_0(980)$:
\begin{equation}
A_{\mu\nu}= e^2 A_{f_0\to\gamma\gamma}(q^2)\; g^{\perp\perp}_{\mu\nu}
\label{4.1}
\end{equation}
where $g^{\perp\perp}_{\mu\nu} q'_\nu =0$ and $g^{\perp\perp}_{\mu\nu}
q_\mu =0$.

The invariant amplitude reads:
\begin{equation}
A_{f_0 \to \gamma \gamma} (q^2)
=\sqrt{N_c} \int \limits_{4m^2}^\infty \frac{dsds'}{\pi^2}
\psi_{f_0}(s)\psi_{\gamma}(s') \frac
{\theta\left (ss'Q^2-m^2\lambda(s,s',Q^2)\right )}
{16\sqrt{\lambda(s,s',Q^2)}}
Z^{(q\bar q)}_{f_0 \to \gamma \gamma}
S^{(tr)}_{f_0 \to \gamma \gamma}(s,s',Q^2) \ ,
\label{4.2}
\end{equation}
where $N_c=3$ is number of colours, $Z^{(n\bar n)}_{f_0 \to \gamma
\gamma} =5\sqrt 2 /9 $ and
$Z^{(s\bar s)}_{f_0 \to \gamma \gamma} = 2 /9 $. The spin factor
$S^{(tr)}_{f_0 \to \gamma \gamma}(s,s',Q^2)$ is similar to that
for $\phi \to \gamma f_0$, namely:
\begin{equation}
S^{(tr)}_{f_0 \to \gamma \gamma}(s,s',q^2)=-2m\left (-s+s'+4m^2
-\frac{4ss'}{s+s'-q^2}\alpha(s,s',q^2) \right ) .
\label{4.3}
\end{equation}
In the limit $Q^2 \to 0$, the term which makes different
$S^{(tr)}_{f_0 \to \gamma \gamma}(s,s',q^2)$ and
$ S^{(tr)}_{\phi \to \gamma f_0}(s,s',q^2)$ vanishes in the integral
(\ref{4.2}), and we have for $A_{f_0 \to \gamma \gamma} (0)$
an expression similar to (\ref{2.14}):
\begin{equation}
A_{f_0 \to \gamma \gamma} (0)
=\frac{m\sqrt{N_c} Z^{(q\bar q)}_{f_0 \to \gamma \gamma} }{4\pi}
\int \limits_{4m^2}^\infty \frac{ds}{\pi}
\psi_{f_0}(s)\psi_{\gamma}(s)
 \left [\sqrt{s(s-4m^2)} -2m^2
\ln \frac{\sqrt{s}+\sqrt{s-4m^2}}{\sqrt{s}-\sqrt{s-4m^2}} \right ]\ .
\label{4.4}
\end{equation}
The photon wave function was found in the analysis of the transition
form factors $\pi^0\to \gamma (Q^2) \gamma$, $\eta \to \gamma (Q^2)
\gamma$, and $\eta' \to \gamma (Q^2) \gamma$ \cite{AMN}: it is shown
in Fig. 5. With this wave
function we calculate $A^{n\bar n}_{f_0 \to \gamma \gamma} (0)$ and
$A^{s\bar s}_{f_0 \to \gamma \gamma} (0)$; these amplitudes plotted
versus $R^2_{f_0(980)}$ are shown in Fig. 2b.

The partial width $\Gamma_{f_0 \to \gamma \gamma}$ is equal to:
\begin{equation}
M_f \Gamma_{f_0 \to \gamma \gamma}=\pi \alpha ^2
|A_{f_0 \to \gamma \gamma} (0) |^2 ,
\label{4.5}
\end{equation}
where
\begin{equation}
A_{f_0 \to \gamma \gamma} (0)=
\cos \varphi A^{(n\bar n)}_{f_0 \to \gamma \gamma} (0) +
\sin \varphi A^{(s\bar s)}_{f_0 \to \gamma \gamma} (0)\; .
\end{equation}
Figure 6 demonstrates the comparison of the calculated
$\Gamma_{f_0 \to \gamma \gamma}$, at different $R^2_{f_0(980)}$ and
$\varphi$, with the data \cite{Pennington}.
It is possible to describe the experimental data ($\Gamma_{f_0(980)
\to\gamma\gamma}=0.28^{+0.09}_{-0.13}$ \cite{Pennington}) at positive
mixing angles as well as at negative ones:
$77^\circ \le \varphi \le 93^\circ$ and $(-54^\circ) \le \varphi \le
(-38^\circ)$. The use of the radius restriction (2) makes the allowed
interval of mixing angles slightly narrower, see (3).

\section{Radiative decays
$\phi(1020) \to \gamma\eta, \gamma\eta', \gamma\pi^0,
\gamma \lowercase{a}_0(980)$}

The decays $\phi(1020) \to \gamma\eta, \gamma\eta', \gamma\pi^0,
\gamma a_0(980)$ do
not provide us with direct information on the quark content of
$f_0(980)$; still, it looks necessary to perform calculations and
comparison with data in order to check the reliability of the method. In
addition, the decay $\phi(1020) \to \gamma\pi^0$ allows us to evaluate
the admixture of the $n\bar n$ component in the $\phi$ meson; as
we saw in Section 2, this admixture affects significantly the value
$\Gamma_{\phi\to\gamma f_0(980)}$.

The amplitude for the transition $\phi \to \gamma P$, where
$P=\eta , \eta' , \pi^0$, has the following structure:
\begin{equation}
A_{\mu\nu}=eA_{\phi \to \gamma P}(q^2)\epsilon_{\mu\nu\alpha\beta}
p_\alpha q_{\beta}\ .
\end{equation}
Radiative decay amplitudes
$\phi \to \gamma\eta, \gamma\eta', \gamma\pi^0$ may be calculated
similarly to the decay amplitude $\phi \to \gamma f_0(980)$,
see Eq. (\ref{2.14}), with
necessary substitutions of the wave functions $\psi_{f_0} \to
\psi_{\eta},\psi_{\eta'},\psi_{\pi}$ as well as charge and spin
factors:
\begin{equation}
A_{\phi \to \gamma P} (0)
=\frac{mZ^{(q\bar q)}_{\phi \to \gamma P}}{4\pi}
\int \limits_{4m^2}^\infty \frac{ds}{\pi}
\psi_\phi(s)\psi_P(s)
\ln \frac{\sqrt{s}+\sqrt{s-4m^2}}{\sqrt{s}-\sqrt{s-4m^2}} \ .
\end{equation}
The charge factors for the considered radiative decays are as
follows: for the $s\bar s$
component in the reactions $\phi \to \gamma\eta, \gamma\eta'$,
$Z^{(s\bar s)}_{\phi\to\gamma\eta}=Z^{(s\bar s)}_{\phi\to\gamma\eta'}
=-2/3$, and for the $\pi^0$ production, $ Z_{\phi \to\gamma\pi^0} = 1$.

For the transitions $\phi \to \gamma\eta$ and $\phi \to \gamma\eta'$
we take into account the dominant $s\bar s$ component only:
$-\sin\theta \, s\bar s$ in $\eta$-meson and
$\cos\theta \, s\bar s$ in $\eta'$-meson, and $\sin\theta =0.6$.

The spin factors are equal to
\begin{equation}
S^{(tr)}_{\phi\to\eta}(s,s',q^2)=S_{\phi\to\eta'}(s,s',q^2)=4m_s,
\qquad
S^{(tr)}_{\phi\to\pi}(s,s',q^2)=4m\ .
\end{equation}
For pion wave function we have chosen $b_\pi=2.0$ GeV$^{-2}$ that
corresponds to $R^2_\pi=10.1$ GeV$^{-2}$, the same radius is fixed
for the $n\bar n$ component in $\eta$ and $\eta'$.
As to the strange component in $\eta$ and $\eta'$, its slope is
the same: $b_{\eta(s\bar s)}=2$ GeV$^{-2}$ that leads to a smaller
radius, $R^2(s\bar s)=8.3$ GeV$^{-2}$.

The results of the calculations versus
branching ratios given by PDG-compilation
\cite{PDG-00} are as follows:
\begin{eqnarray}
&&BR(\phi\to \eta\gamma)=1.46\cdot 10^{-2}\ , \qquad
BR_{PDG}(\phi\to \eta\gamma)=(1.30\pm 0.03)\cdot 10^{-2}\ ,
\nonumber \\
&&BR(\phi\to \eta'\gamma)=0.97\cdot 10^{-4}\ , \qquad
BR_{PDG}(\phi\to \eta'\gamma)=(0.67 ^{+ 0.35}_{-0.31})\cdot 10^{-4}\ .
\end{eqnarray}
The calculated branching ratios agree reasonably with those given in
\cite{PDG-00}.

For the process $\phi \to \gamma \pi^0$ the compilation \cite{PDG-00}
gives $BR(\phi\to\gamma \pi^0 )=(1.26\pm 0.10)\cdot 10^{-3}$, and this
value requires $|\sin\varphi_V|\simeq 0.07$
(or $|\varphi_V|\simeq 4^\circ$),
for with just this
admixture of the $n\bar n$ component in $\phi (1020)$ we reach the
agreement with data.
 However, in the estimation of the allowed regions for the mixing angle
$\varphi$ (see Fig. 3), we use
\begin{equation}
|\varphi_V| = 4^\circ\pm 4^\circ
\end{equation}
considering the accuracy inherent to the quark model to be
comparable with the obtained small value of $|\varphi_V|$.

The process
$\phi(1020) \to \gamma a_0(980)$ depends also on the mixing angle
$|\varphi_V|$: the decay amplitude is proportional to $\sin \varphi_V$,
namely, $A_{\phi \to \gamma a_0} =
\sin \varphi_V A^{n\bar n}_{\phi \to \gamma a_0} $. The
amplitude $A^{n\bar n}_{\phi \to \gamma a_0}$ is equal to that
for the process $\phi \to \gamma f_0$, up to a numerical factor:
$A^{n\bar n}_{\phi \to \gamma a_0}/A^{n\bar n}_{\phi \to \gamma f_0} =
Z^{n\bar n}_{\phi \to \gamma a_0} /Z^{n\bar n}_{\phi \to \gamma f_0}
=3$ because $Z^{n\bar n}_{\phi \to \gamma a_0} =1$ (the value
$3A^{n\bar n}_{\phi \to \gamma f_0}$ is shown on Fig. 2a). We have for
the region $R^2_{a_0(980)} \sim 8$ GeV$^{-2}$ - $12$ GeV$^{-2}$:
\begin{equation}
BR\left (\phi (1020) \to \gamma a_0 (980) \right )\simeq
\sin^2 \varphi_V \cdot (14\pm 3) \cdot 10^{-4}
\end{equation}
with lower values for $R^2_{a_0(980)} \sim 8$ GeV$^{-2}$ and larger ones
for $R^2_{a_0(980)} \sim 12$ GeV$^{-2}$. At $\sin^2 \varphi_V \leq 0.02$,
we predict $BR(\phi (1020) \to \gamma a_0 (980) )\leq 0.28 \cdot
10^{-4}$.

In recent paper \cite{machasov} the partial width for
$\phi (1020) \to \gamma \eta\pi^0$ was measured, it was found
$BR(\phi (1020) \to \gamma \eta\pi^0;\; M_{\eta\pi} > 900$ MeV$)=
(0.46\pm0.13)\cdot 10^{-4}$. If at $ M_{\eta\pi} > 900$ MeV the
ratio of the resonance/background is of the order of the unity,
that is quite possible, the value found in \cite{machasov} is in
agreement with small value  of $|\varphi_V| $.

\section{Conclusion}
Figure 7 demonstrates the $(\varphi , R^2_{f_0(980)})$-plot
where the allowed areas for the reactions $\phi (1020) \to \gamma f_0
980)$ and $\phi (1020) \to \gamma \gamma$ are shown. We see that
the radiative decays
$\phi(1020) \to\gamma f_0(980)$ and $f_0(980) \to \gamma\gamma$
are well described in the framework of the hypothesis of the
dominant $q\bar q$
structure of $f_0(980)$.
For flavour wave  function determined by $\Psi_{flavour}(f_0(980))=
n\bar n\cos \varphi +s\bar s\sin \varphi$, the solutions found are
either $\varphi=-48^\circ\pm 6^\circ$ or $\varphi=85^\circ\pm 5^\circ$.
The solution with negative $\varphi$ seems
more preferable, that means the $q\bar q$
component is rather close to the flavour octet
($\varphi_{octet}=-54.7^\circ$). Such a proximity to the octet may be
related to the following scenario:
 the broad resonance $f_0(1530^{+90}_{-250})$, which is,
according to the $K$-matrix analysis,  the
descendant of the lightest scalar glueball after accumulation of widths
of neighbouring resonances took the singlet component from a
predecessor of $f_0(980)$, thus turning $f_0(980)$ into an almost pure
octet.

The dominance of  $q\bar q$ component does not exclude the
existence of the other components in $f_0(980)$. The location of the
resonance pole near the $K\bar K$ threshold definitely points on a
certain admixture of the long-range $K\bar K$ component in $f_0(980)$.
To investigate this admixture the precise measurements of the $K\bar
K$ spectra in the interval $1000- 1150$ MeV are necessary: only these
spectra could shed the light on the role of the long--range $K\bar K$
component in $f_0(980)$.

The existance of the long-range $K\bar K$ component or that of gluonium
in the $f_0(980)$ results in a decrease of the $s\bar s$ fraction in
the $q\bar q$ component: for example, if the long-range $K\bar K$ (or
gluonium) admixture is of the order of 15\%, the data require either
$\varphi=- 45^\circ \pm 6^\circ$ or $\varphi=83^\circ \pm 4^\circ$.

We thank D. V. Bugg, L. G. Dakhno, V. S. Fadin, L. Montanet and
A. V. Sarantsev for useful discussions and illuminative information.
The work is supported by the RFBR grant 01-02-17861.

\newpage

\begin{figure}
\centerline{\epsfig{file=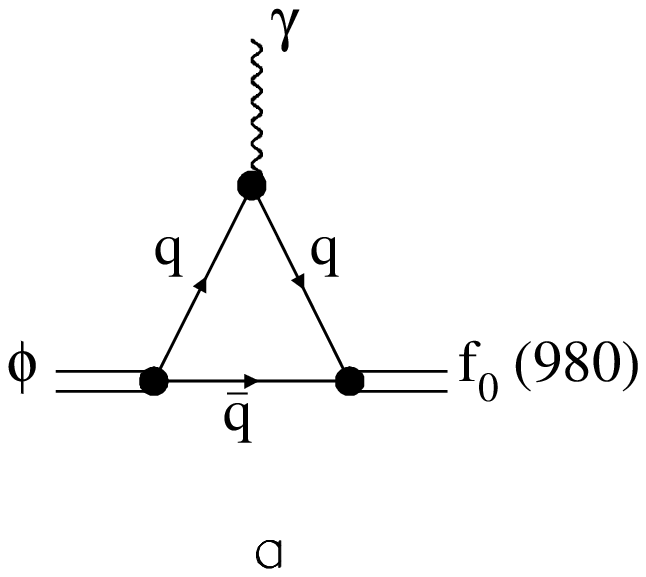,height=6cm}\hspace{1cm}
            \epsfig{file=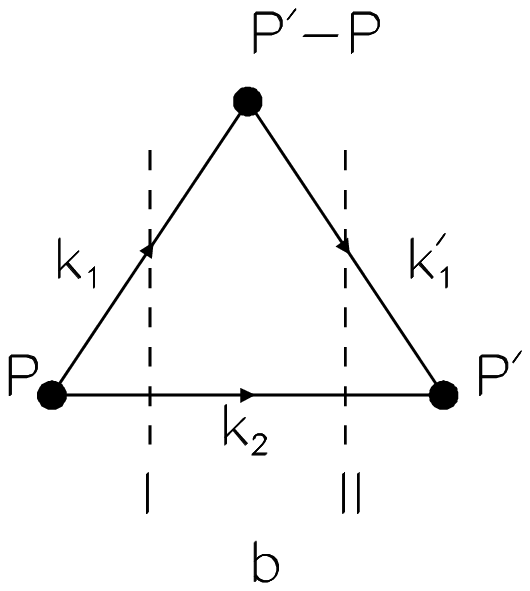,height=6cm}}
\caption{a) Diagrammatic representation of the transition
$\phi(1020)\to\gamma f_0(980)$. b) Three-point quark diagram: dashed
lines I and II mark the two cuttings in the double spectral
representation. }
\end{figure}

\begin{figure}
\centerline{\epsfig{file=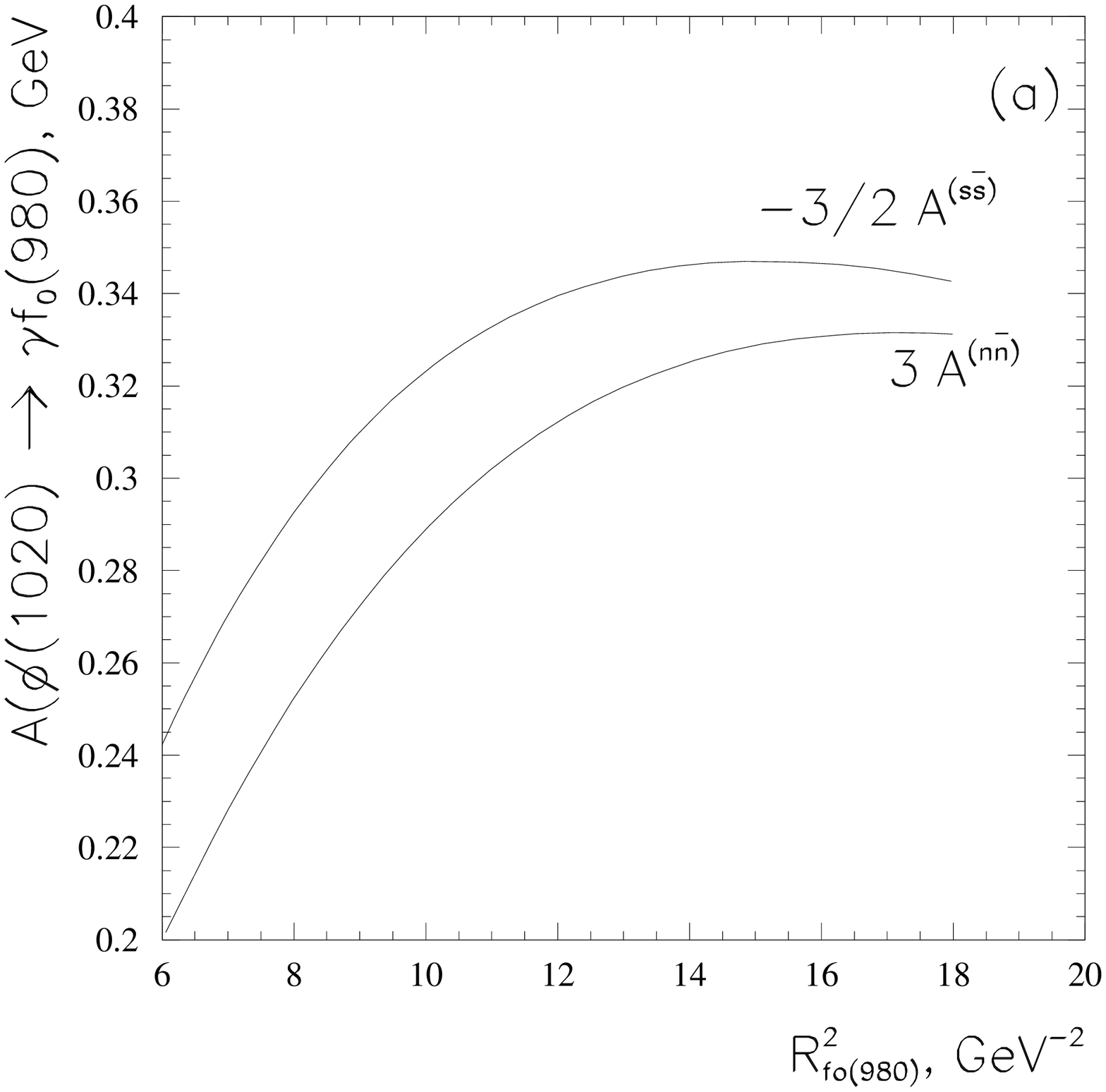,width=8cm}\hspace{0.5cm}
            \epsfig{file=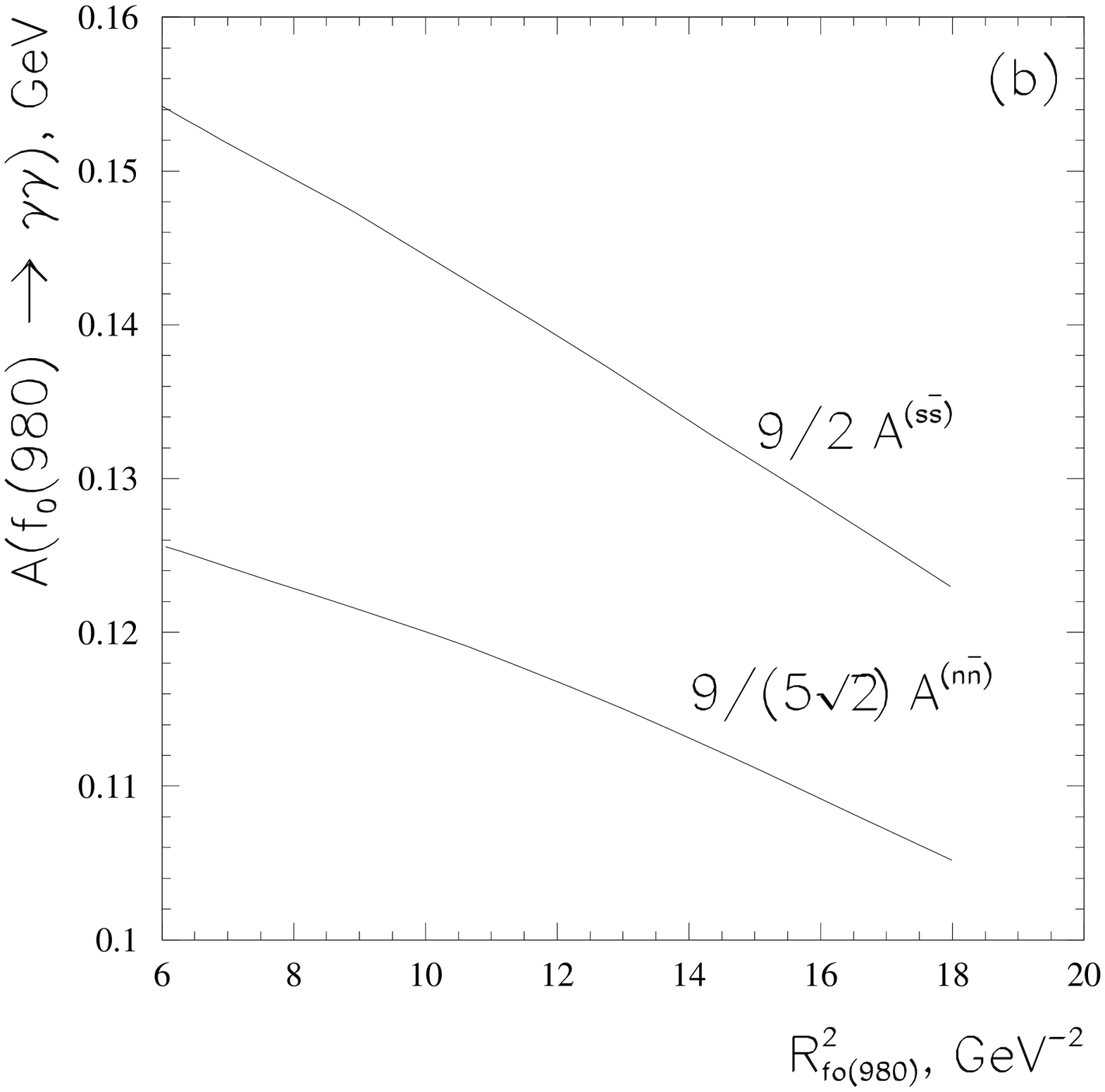,width=8cm}}
\caption{Amplitudes for strange and non-strange components,
$s\bar s$ and $n\bar n$, as functions of the
$f_0(980)$-meson radius squared: a)
$A^{(n\bar n)}_{\phi\to\gamma f_0}(0)/
Z^{(n\bar n)}_{\phi\to\gamma f_0}$ and
$A^{(s\bar s)}_{\phi\to\gamma f_0}(0)/
Z^{(s\bar s)}_{\phi\to\gamma f_0}$,
b) $A^{(n\bar n)}_{\phi\to\gamma\gamma}(0)/
Z^{(n\bar n)}_{\phi\to\gamma\gamma}$
and $A^{(s\bar s)}_{\phi\to\gamma\gamma}(0)/
Z^{(s\bar s)}_{\phi\to\gamma\gamma}$.}
\end{figure}

\begin{figure}
\centerline{\epsfig{file=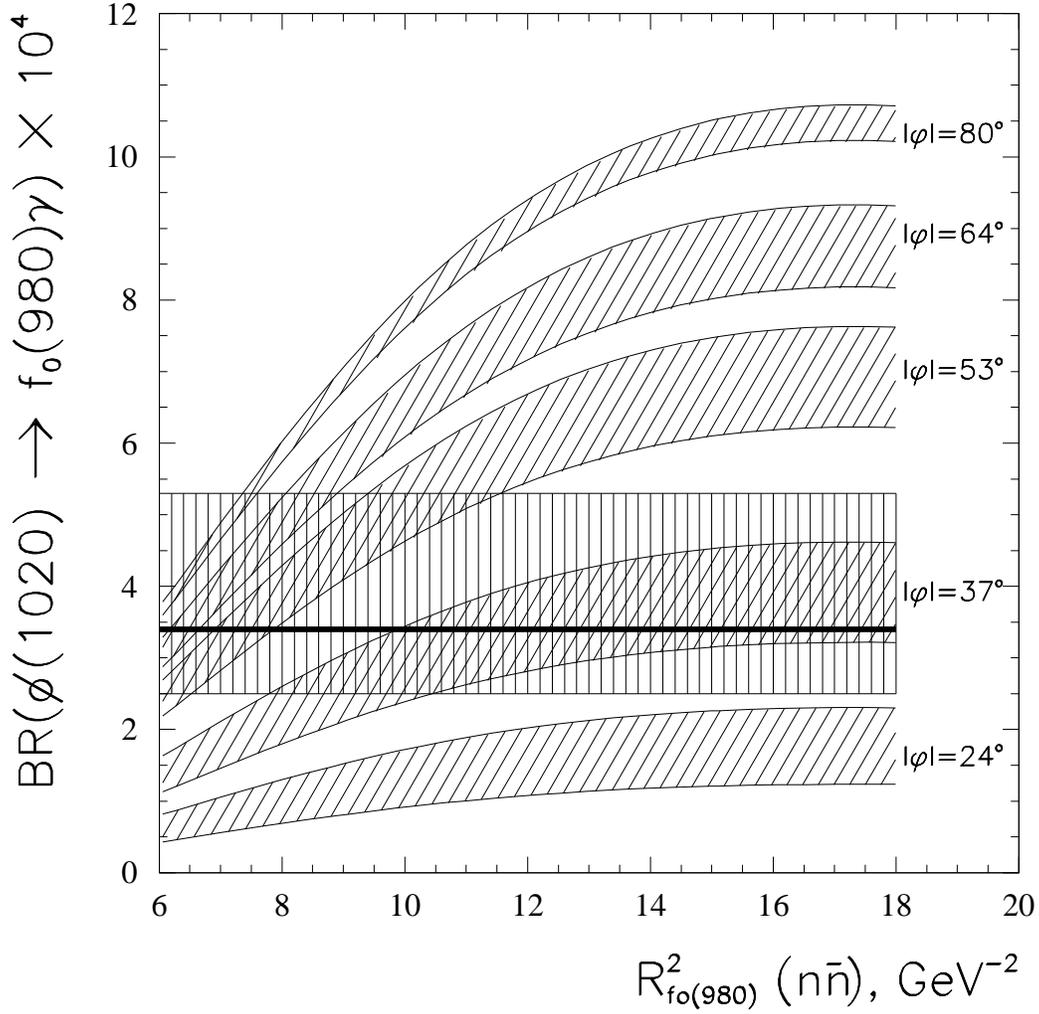,width=15cm}}
\caption{Branching ratio $BR(\phi(1020)\to\gamma f_0(980))$ as a
function of the radius squared of the $n\bar n$ component in
$f_0(980)$. The band with vertical shading stands for the experimental
magnitude; five other bands, with skew shading, correspond to
$|\varphi|=24^\circ,37^\circ, 53^\circ, 64^\circ, 80^\circ$ at
$-8^\circ\le\varphi_V\le 8^\circ $.}
\end{figure}

\begin{figure}
\centerline{\epsfig{file=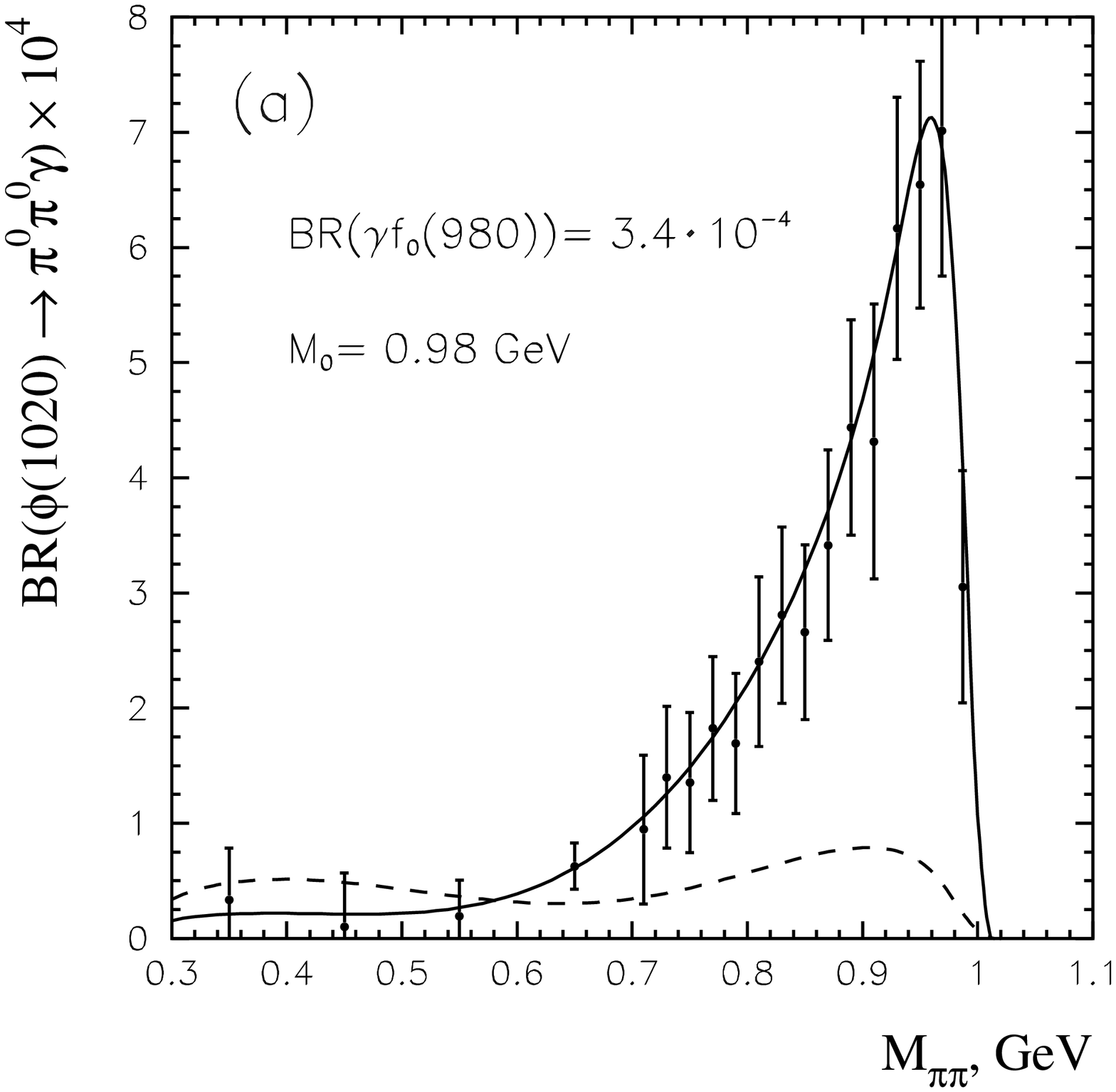,height=9cm}\hspace{-1cm}
            \epsfig{file=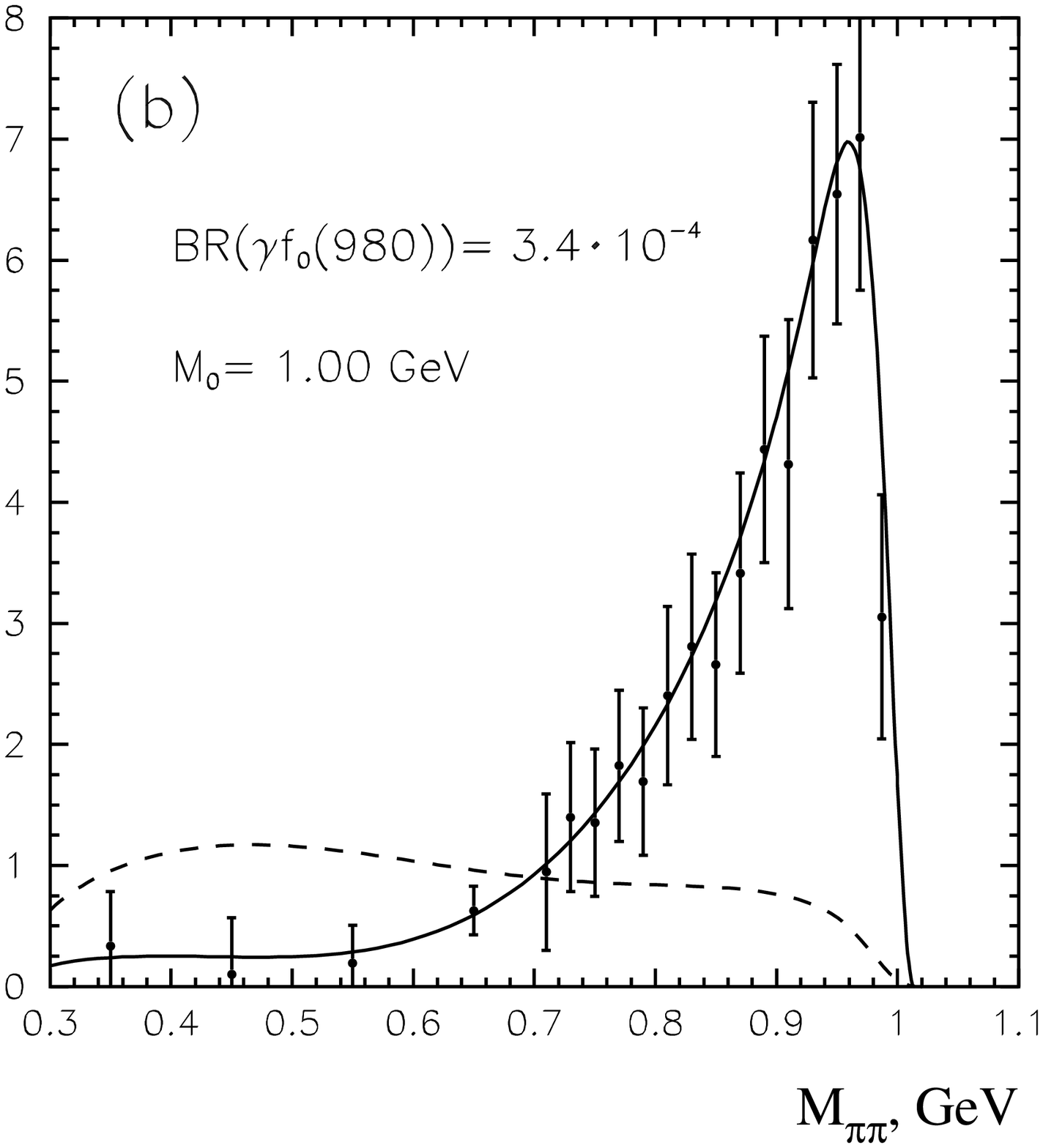,height=9cm}}
\vspace{-1cm}
\centerline{\epsfig{file=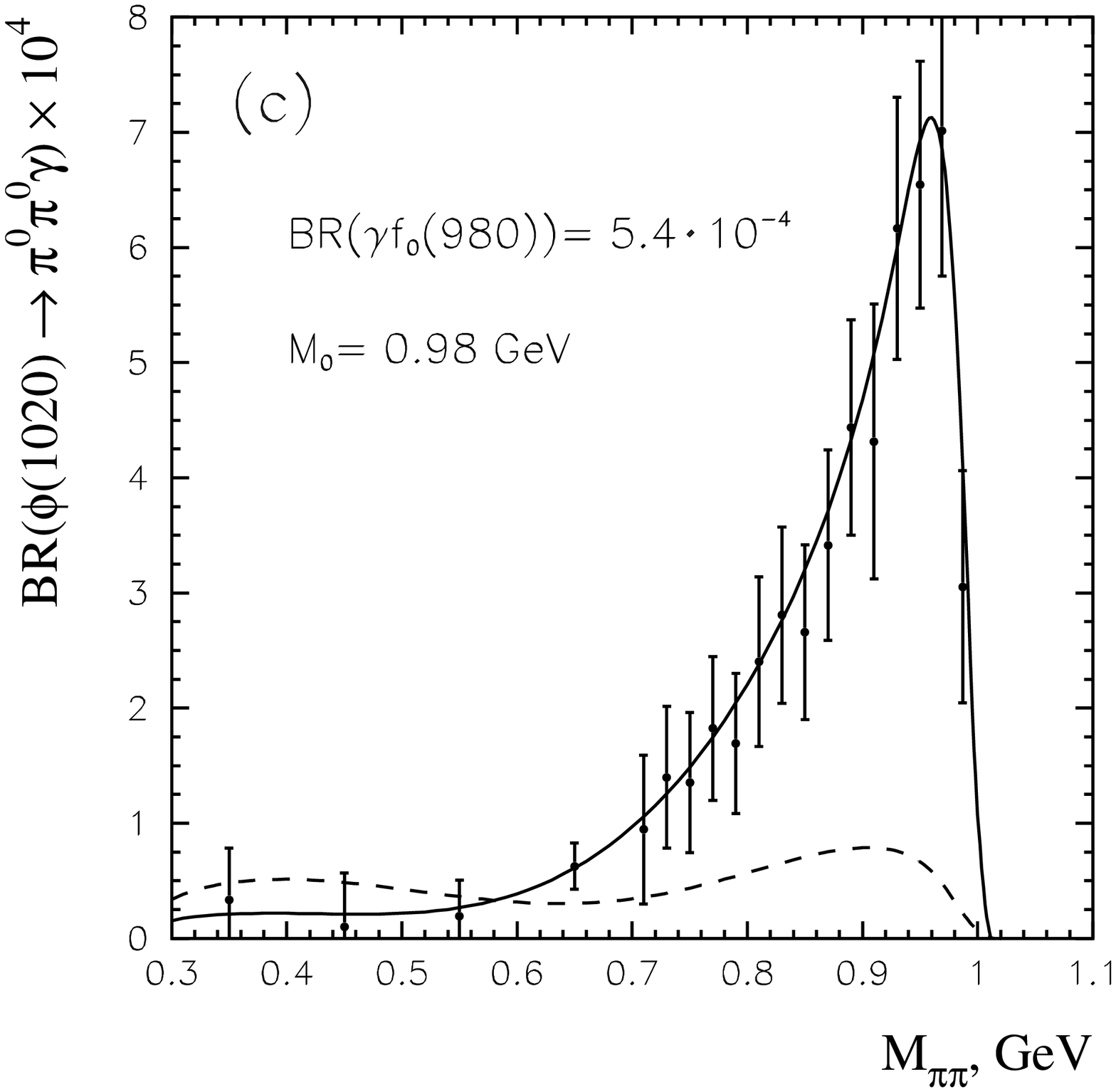,height=9cm}\hspace{-1cm}
            \epsfig{file=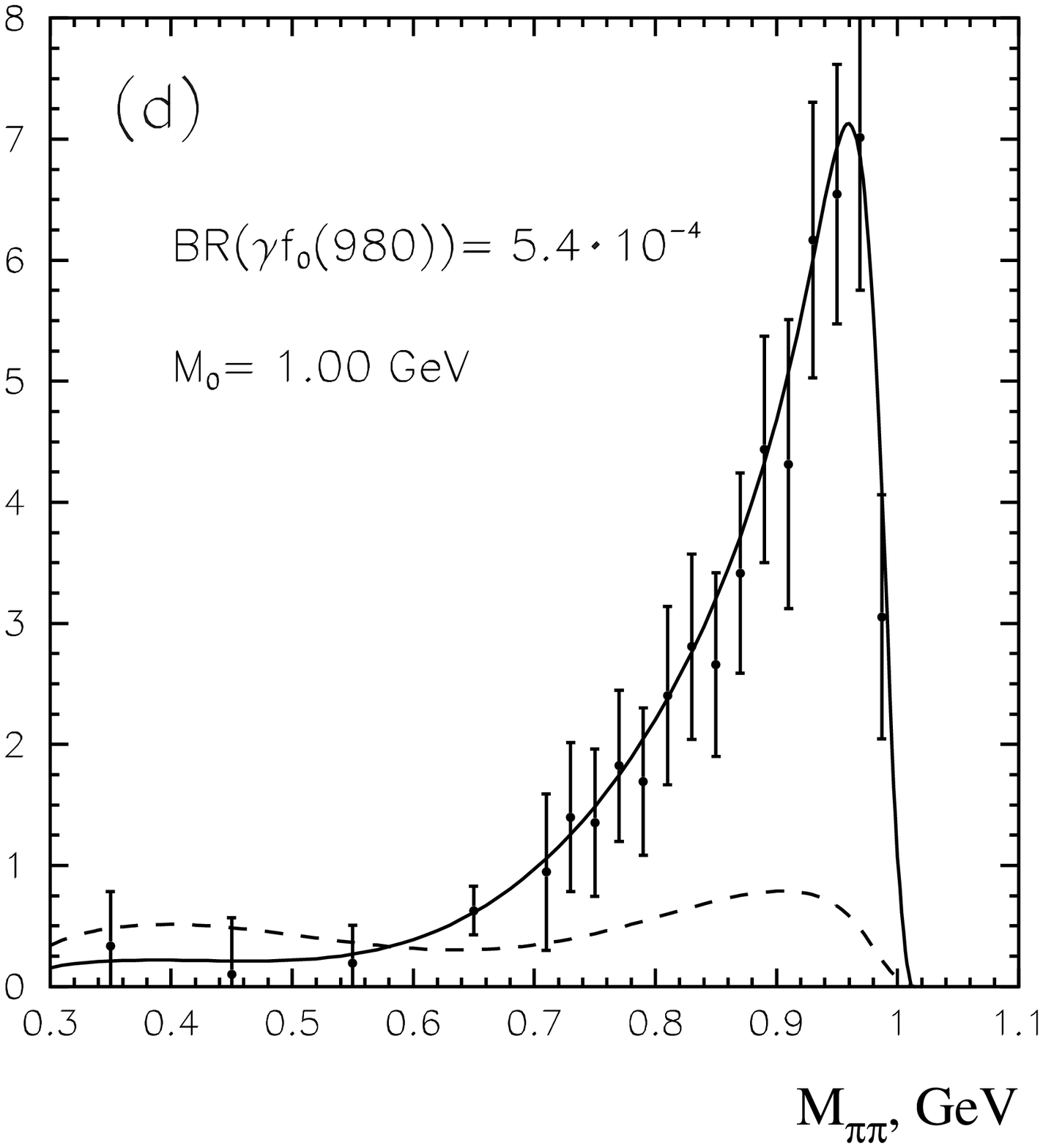,height=9cm}}
\caption{Branching ratios $BR(\phi(1020) \to \pi^0\pi^0\gamma)$
calculated with the Flatt\'e formula (26) and  parametrization
given by (28) (solid curves) at different $BR(\phi(1020) \to \gamma
f_0(980))$ and $M_{f_0(980)}$. Dashed line shows the background
contribution which corresponds to the elimination of the resonance term
in (39). The data are taken from [7].}
\end{figure}

\begin{figure}
\centerline{\epsfig{file=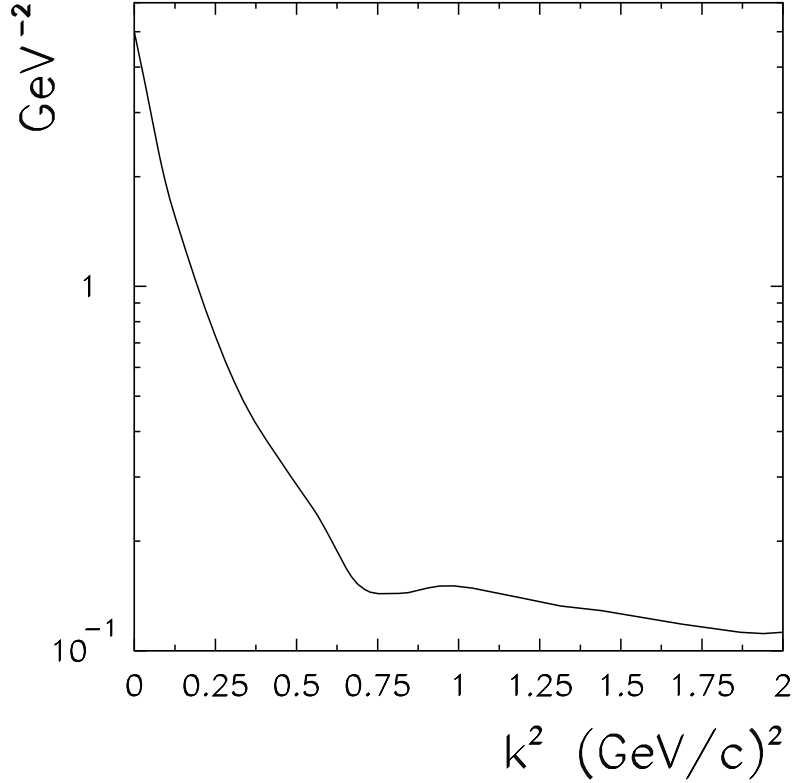,width=8cm}}
\caption{Photon wave function for non-strange quarks,
$\psi_{\gamma\to n\bar n}(k^2)=g_\gamma(k^2)/(k^2+m^2)$, where
$k^2=s/4-m^2$; the wave function for the $s\bar s$
component is equal to
$\psi_{\gamma\to s\bar s}(k^2)=g_\gamma(k^2)/(k^2+m^2_s)$; the
constituent quark masses are  $m$=350 MeV and $m_s$=500 MeV.}
\end{figure}

\begin{figure}
\centerline{\epsfig{file=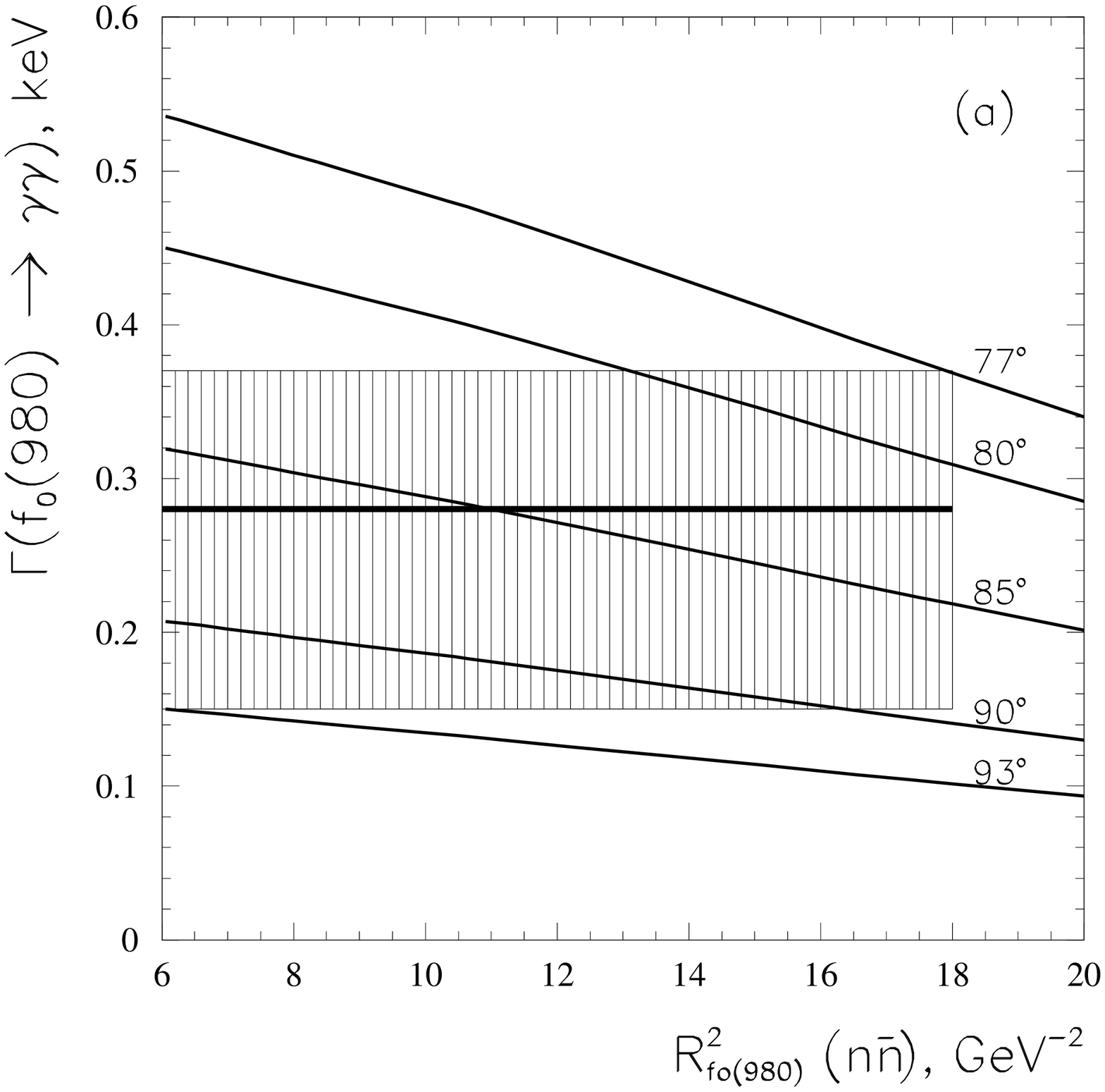,width=8cm}\hspace{1cm}
            \epsfig{file=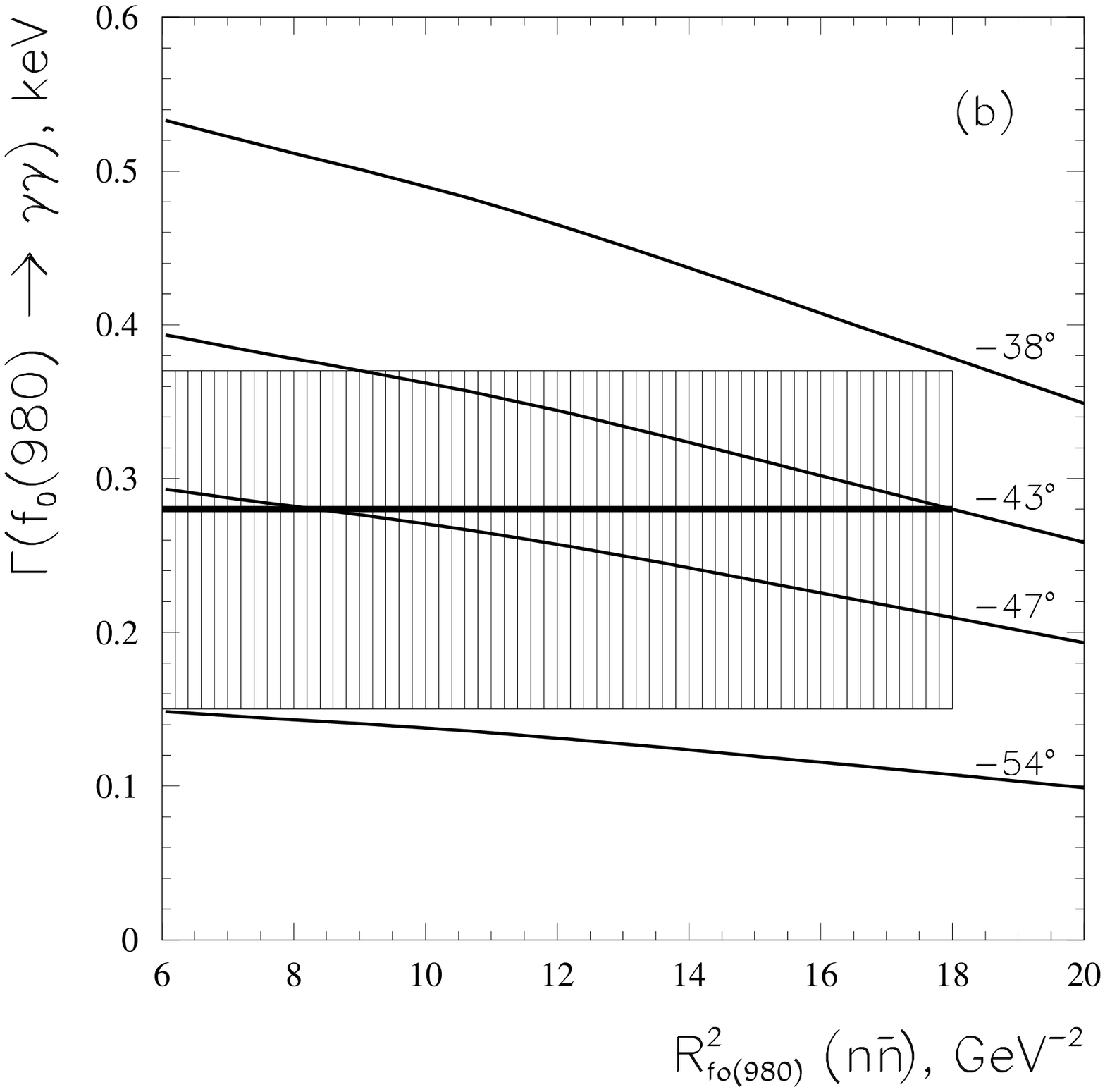,width=8cm}}
\caption{Partial width $\Gamma_{f_0(980) \to \gamma\gamma}$;
experimental data are from [22] (shaded area). a) Curves are
calculated for positive mixing angles
$\varphi=77^\circ,80^\circ,85^\circ,90^\circ,93^\circ$ and b) negative angles
$\varphi=-38^\circ,-43^\circ,-47^\circ,-54^\circ$.}
\end{figure}

\begin{figure}
\centerline{\epsfig{file=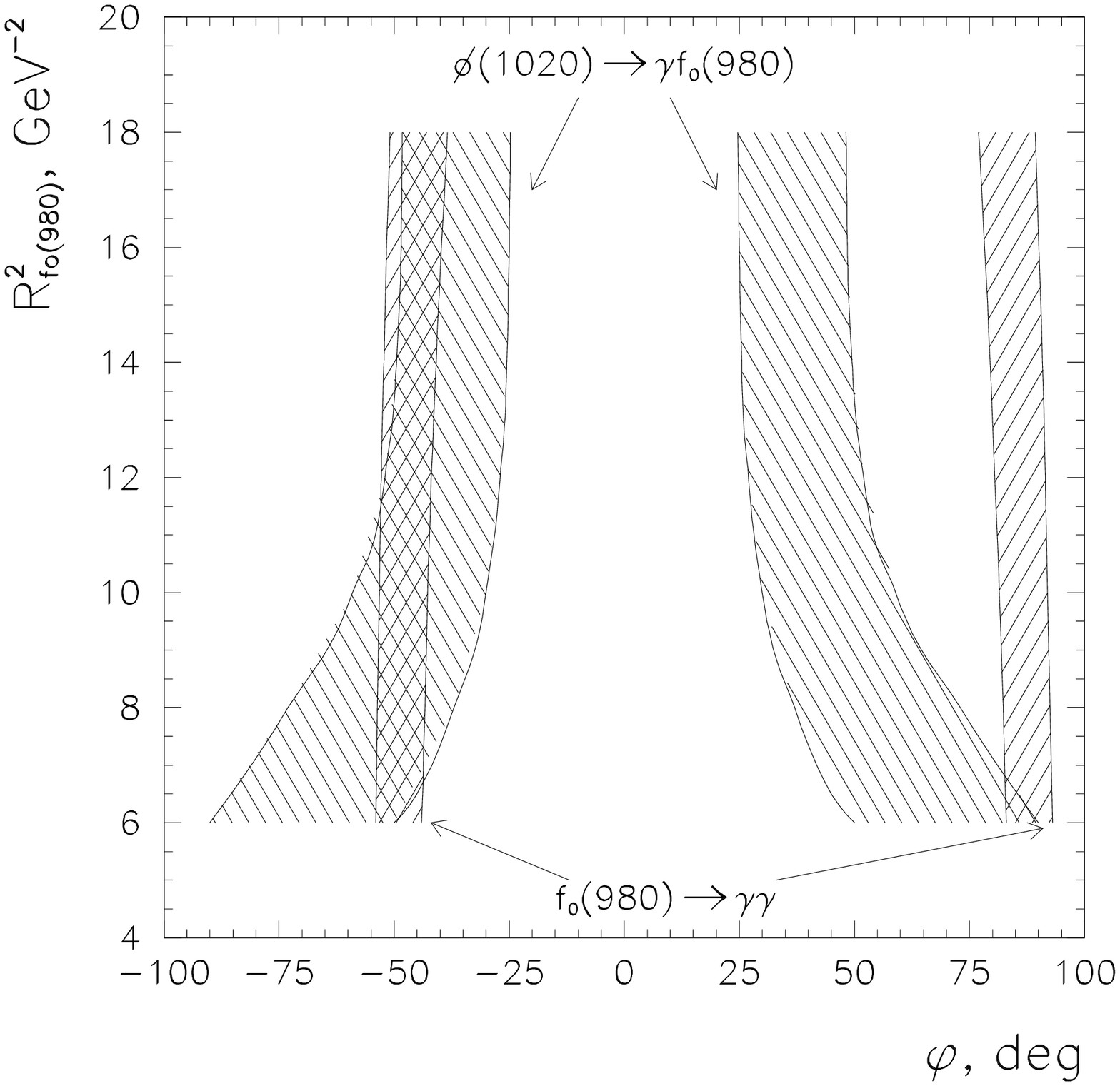,width=15cm}}
\caption{The $(\varphi,R^2_{f_0(980)})$-plot: the shaded areas are
the allowed ones for the reactions
$\phi(1020)\to\gamma f_0(980)$ and $ f_0(980) \to \gamma\gamma$.}
\end{figure}


\begin{thebibliography}{fussy}
\bibitem{klempt}E. Klempt, {\it Meson spectroscopy: glueballs, hybrids,
and $q$ anti-$q$ mesons}, hep-ex/0101031 (2001).
\bibitem{petry} R. Ricken, M. Koll, D. Merten, B.C. Metsch and
H.R. Petry, Eur. Phys. J. A {\bf 9}, 221 (2000); hep-ph/0008221.
\bibitem{1} L. Montanet, Nucl. Phys. Proc. Suppl. {\bf 86}, 381 (2000).
\bibitem{3} V. V. Anisovich, Physics-Uspekhi, {\bf 41} 419 (1998).
\bibitem{Bugg} B. S. Zou and D. V. Bugg, Phys. Rev. D {\bf 50}, 591
(1994).
\bibitem{5} D. Morgan and M. R. Pennington,
Phys. Rev. D {\bf 48}, 1185 (1993).
\bibitem{6} CMD-2 Collaboration: R. R. Akhmetshin {\it et al.}, Phys.
Lett. B {\bf 462}, 371 (1999); {\bf 462}, 380 (1999).
\bibitem{7} SND Collaboration: M. N. Achasov {\it et al.},
Phys. Lett. B {\bf 485}, 349 (2000).
\bibitem{8}
V. V. Anisovich, Yu. D. Prokoshkin and A. V. Sarantsev,
Phys. Lett. B {\bf 389}, 388 (1996).
\bibitem{aanis} A. V. Anisovich and A. V. Sarantsev, Phys. Lett. B
{\bf 413}, 137 (1997).
\bibitem{lattice}
G. S. Bali et al., Phys. Lett. B {\bf 309}, 378 (1993); \\
J. Sexton, A. Vaccarino and D. Weingarten, Phys. Rev. Lett.
{\bf 75}, 4563 (1995);\\
C. J. Morningstar and M. Peardon, Phys. Rev. D {\bf 56}, 4043 (1997).
\bibitem{accum} V. V. Anisovich, D. V. Bugg and A. V. Sarantsev,
Phys. Rev. D {\bf 58}:111503 (1998).
\bibitem{okun}I. Yu. Kobzarev, N. N. Nikolaev and L. B. Okun,
Sov. J. Nucl. Phys. {\bf 10}, 499 (1970);\\
L. Stodolsky, Phys. Rev. D {\bf 1}, 2683 (1970);\\
I. S. Shapiro, Nucl. Phys. A {\bf 122}, 645 (1968).
\bibitem{t'Hooft} G. t'Hooft, Nucl. Phys. B {\bf 72}, 161 (1974);\\
G. Veneziano, Nucl. Phys. B {\bf 117}, 519 (1976).
\bibitem{Anis_zeit}
A. V. Anisovich, V. V. Anisovich and A. V. Sarantsev,
Zeit. Phys. A {\bf 359}, 173 (1997).
\bibitem{PDG-00}PDG Group, D. E. Groom {\it et al.},
Eur. Phys. J. C {\bf 15}, 1 (2000).
\bibitem{torn} S. Spanier, N.A. T\"ornqvist,
Eur. Phys. J. C {\bf 15}, 437 (2000).
\bibitem{AMN} V. V. Anisovich, D. I. Melikhov and V. A. Nikonov,
Phys. Rev. D {\bf 55}, 2918 (1997); {\bf 52}, 5295 (1995).
\bibitem{AKMS}
V. V. Anisovich, M. N. Kobrinsky, D. I. Melikhov and
A. V. Sarantsev, Nucl. Phys. A {\bf 544}, 747 (1992).
\bibitem{melikhov}D. I. Melikhov
Phys. Rev. D {\bf 56}, 7089 (1997).
\bibitem{melikhov-s}
D. I. Melikhov and B. Stech, Phys. Rev. D {\bf 62}:014006 (2000).
\bibitem{GAMS}
Yu. D. Prokoshkin {\it et al.}, Physics-Doklady {\bf 342},
473 (1995);\\
D. Alde {\it et al.}, Z. Phys. C {\bf 66}, 375 (1995).
\bibitem{radii}
V. V.Anisovich, D. V. Bugg and A. V. Sarantsev, Phys. Lett. B
{\bf 437}, 209 (1998); Phys. Atom. Nucl. {\bf 62 }, 289 (1999).
\bibitem{Flatte} S. M. Flatt\'e , Phys. Lett. B {\bf 63}, 224 (1976).
\bibitem{Andrei}
D. V. Bugg, A. V. Sarantsev and B. S. Zou, Nucl. Phys. B {\bf 471},
59 (1996);\\
A. V. Sarantsev, private communication.
\bibitem{f-gg}
A. V.Anisovich, V. V.Anisovich, D. V. Bugg and V.A. Nikonov,
Phys. Lett. B {\bf 456}, 80 (1999).
\bibitem{AA}
A. V. Anisovich and V. V. Anisovich, Phys. Lett. B {\bf 467},
289 (1999).
\bibitem{Pennington} M. Boglione and M. R. Pennington,
Eur. Phys. J. C {\bf 9}, 11 (1999).
\bibitem{PDG-98}PDG Group, C. Caso {\it et al.},
Eur. Phys. J. C {\bf 3}, 1 (1998).
\bibitem{eta} A.V. Anisovich, V.V. Anisovich, L. Montanet and V.A.
Nikonov, Eur. Phys. J. A {\bf 6}, 247 (1999).
\bibitem{machasov} M. N. Achasov et al. Phys. Lett. {\bf B479}, 53
(2000).

\end{thebibliography}
\end{document}